\documentclass[aps,pra,twocolumn,amsmath,amssymb,groupedaddress,showpacs]{revtex4}
\usepackage{graphicx}
\usepackage{dcolumn}
\usepackage{CJK}
\usepackage{bm}

\begin{document}


\title{Bloch-Landau-Zener dynamics in single-particle Wannier-Zeeman systems}

\author{Yongguan Ke$^{1,2}$}
\author{Xizhou Qin$^{1,2}$}
\author{Honghua Zhong$^{1,2}$}
\author{Jiahao Huang$^{1,2}$}
\author{Chunshan He$^{1}$}

\author{Chaohong Lee$^{1,2}$}
\altaffiliation{Corresponding author. Email: chleecn@gmail.com}

\affiliation{$^{1}$State Key Laboratory of Optoelectronic Materials and Technologies, School of Physics and Engineering, Sun Yat-Sen University, Guangzhou 510275, China}
\affiliation{$^{2}$Institute of Astronomy and Space Science, Sun Yat-Sen University, Guangzhou 510275, China}

\date{\today}

\begin{abstract}
Stimulated by the experimental realization of spin-dependent tunneling via gradient magnetic field [Phys. Rev. Lett. 111, 225301 (2013); Phys. Rev. Lett. 111, 185301 (2013)], we investigate dynamics of Bloch oscillations and Landau-Zener tunneling of single spin-half particles in a periodic potential under the influence of a spin-dependent constant force.
In analogy to the Wannier-Stark system, we call our system as the Wannier-Zeeman system.
If there is no coupling between the two spin states, the system can be described by two crossing Wannier-Stark ladders with opposite tilts.
The spatial crossing between two Wannier-Stark ladders becomes a spatial anti-crossing if the two spin states are coupled by external fields.
For a wave-packet away from the spatial anti-crossing, due to the spin-dependent constant force, it will undergo spatial Landau-Zener transitions assisted by the intrinsic intra-band Bloch oscillations, which we call the Bloch-Landau-Zener dynamics.
If the inter-spin coupling is sufficiently strong, the system undergoes adiabatic Bloch-Landau-Zener dynamics, in which the spin dynamics follows the local dressed states.
Otherwise, for non-strong inter-spin couplings, the system undergoes non-adiabatic Bloch-Landau-Zener dynamics.
\end{abstract}


\maketitle

\section{introduction}\label{Sec1}

The dynamics of a particle in a periodic potential under the action of an additional constant force is a traditional and fundamental problem~\cite{Nenciu1991, Raizen1997, Korsch2002, Xiao2010}.
Two most elementary quantum transport phenomena in this fundamental problem are intra-band Bloch oscillations~\cite{Bloch1929} and inter-band Landau-Zener (LZ) tunneling~\cite{Zener1934}.
Due to combination of the periodic potential and the constant force, the system consists of a series of equidistant energy levels, now called Wannier-Stark (WS) ladders~\cite{Wannier1960}.
In addition to the traditional systems of electrons in solid crystals~\cite{Mendez1988, Waschke1993,Dekorsy1994,Lyssenko1997}, this fundamental problem has been investigated by using several other systems, such as the systems of photons in waveguide arrays~\cite{Peschel1998,Pertsch1999, Morandotti1999, Pertsch2002}, cold atoms in optical lattices~\cite{Dahan1996, Niu1996, Wilkinson1996, Peik1997, Bharucha1997, Morsch2001}.
In particular, attribute to the well-developed techniques of manipulating and detecting cold atoms in optical lattices, there continuously emerges new research interests on this traditional and fundamental problem~\cite{Atala2013, Kolovsky2014, Clade2014}.
For a weak constant force, the dynamics of Bloch oscillations can be analyzed by using single-band and tight-binding approximations~\cite{Hartmann2004,Holthaus1996}.
However, if the constant force is strong enough, tunneling effects between different Bloch bands could not be neglected and the dynamics should be described with coherent superpositions of intra-band Bloch oscillations and inter-band LZ tunneling~\cite{Breid2006,Breid2007, Dreisow2009}.
Although the WS systems of spinless particles have been studied extensively, there is few report on the WS systems with spin-dependent tilts and inter-spin couplings~\cite{PhysRevA.82.033602}, which is called the Wannier-Zeeman (WZ) systems in this article.

The simplest WZ system, which has a spin-$\frac{1}{2}$ particle in a periodic potential under the influence of a spin-dependent constant force, can be regarded as two crossing WS systems with opposite tilts.
Recently, spin-dependent tunneling and spin-orbit coupling of ultracold atoms in optical lattices have been experimentally realized by applying a gradient magnetic field~\cite{Kennedy2013, Miyake2013, Aidelsburger2013}.
Due to the opposite magnetic dipole moments for the two atomic spin states, their Zeeman energy offsets are also opposite.
In these experiments~\cite{Kennedy2013, Miyake2013, Aidelsburger2013}, the two atomic spin states are not coupled, so the system can be regarded as two independent WS systems of opposite tilts.
In such a WZ system, besides intra-band Bloch oscillations and inter-band LZ tunneling in the same spin component, it is possible to find novel dynamics between two spin components, such as LZ tunneling between two spin states.
To observe such a kind of LZ tunneling between spin states, one has to introduce a coupling between spin states.
It naturally arises two open questions: (i) how the coupling between spin states affects Bloch oscillations? and (ii) how Bloch oscillations affect the LZ tunneling between spin states?

In this article, we study Bloch oscillations and LZ tunneling of a coupled two-level atom in a one-dimensional (1D) optical lattice under the influence of a gradient magnetic field.
By using the experimental techniques for simulating two-dimensional (2D) lattice models with spin-dependent tunneling~\cite{Kennedy2013, Miyake2013, Aidelsburger2013}, our model can be realized by reducing their 2D optical lattices to 1D ones.
If the magnetic field gradient is not very strong, the tunneling between neighboring lattice sites is not suppressed and there is no need to induce additional Raman lasers to assist tunneling between neighboring lattice sites.
The two internal spin states can be coupled by laser fields and the coupling strength can be modulated by adjusting the laser strengths.
The coupling between two internal states opens the crossing of two WS systems corresponding to the two internal states, and so spatial LZ tunneling between internal states may occur when the atom passes through the anti-crossing region.
We explore the novel inter-spin LZ tunneling induced by the intrinsic intra-band Bloch oscillations.
Dependent on the inter-spin coupling strength, the system show adiabatic and non-adiabatic Bloch-Landau-Zener dynamics for strong and non-strong couplings, respectively.
In the adiabatic Bloch-Landau-Zener dynamics, the spin dynamics follows the local dressed states and the wave-packet undergoes joint Bloch-like oscillations of the two spin components.
In the non-adiabatic Bloch-Landau-Zener dynamics, the spin dynamics could not follows the local dressed states and the two spin components undergo separate Bloch oscillations.

This article is constructed as follows.
In Sec.~\ref{Sec1}, we briefly introduce the background and our motivation.
In Sec.~\ref{sec2}, we describe the physical model and simplify the original Hamiltonian under the single-band tight-binding approximation.
In Sec.~\ref{sec3}, we solve the eigenvalue problem and discuss the WZ states and the WZ ladders.
In Sec.~\ref{sec4}, we explore the dynamics of Bloch oscillations and LZ tunneling for different coupling strengths.
In Sec.~\ref{sec5}, we summarize our results and discuss the validity of the single-band tight-binding approximation.

\section{Model}\label{sec2}

We consider a ${}^{87}$Rb atom in a 1D optical lattices.
The atom may occupy two possible hyperfine spin states  $\left|\uparrow\right\rangle\equiv\left|{f=2, m_f=2}\right\rangle$ and $\left|\downarrow\right\rangle\equiv\left|{f=2, m_f=-2}\right\rangle$ (or another pair of suitable hyperfine spin states with opposite magnetic dipole moments) which are coupled by lasers/microwaves.
Similar to the creation of spin-dependent tunneling~\cite{Kennedy2013, Miyake2013, Aidelsburger2013}, a magnetic field gradient is applied to generate the spin-dependent constant force along the optical lattice direction, taking as z direction.
Such a system obeys the following Hamiltonian,
\begin{equation}
\label{equ:origin}
\hat{H}  =  \hat H_0 +\hat H_t + \hat H_c,
\end{equation}
with
\begin{eqnarray}
\hat H_0 =\left(
\begin{matrix}
-{{\hbar}^2 \over {2M}} {\mathrm{d}^2 \over {\mathrm{d}z^2}}+ V(z) & 0 \\
  0   & -{{\hbar}^2 \over {2M}} {\mathrm{d}^2 \over {\mathrm{d}z^2}}+ V(z)
\end{matrix}
\right) \nonumber,
\end{eqnarray}
\begin{eqnarray}
  \hat H_t=
\left(
\begin{matrix}
  F_{m_f}z & 0 \\
  0   & -F_{m_f}z
\end{matrix}
\right), \;\;\;
\hat H_c=
\left(
\begin{matrix}
  0 & \hbar \Omega \over 2 \\
  \hbar \Omega \over 2   & 0
\end{matrix}
\right).\nonumber
\end{eqnarray}
The term ${\hat H_0}$ describes the spin-independent motions, in which, $V(z)={{{V_0}} \over 2}\cos \left({2{\kappa_0}z}\right)$ is a periodic potential formed by standing-wave lasers~\cite{Bloch2008,Morsch2006}, $M$ is the atomic mass, $V_0$ is the potential depth proportional to the laser intensity, and $\kappa_0=2\pi/\lambda$ is the laser wave vector.
Obviously, the periodic potential has a period $d=\lambda/2$ determined by the laser wavelength $\lambda$.
The term ${\hat H}_t$ denotes the spin-dependent tilts induced by the Zeeman shifts of a gradient magnetic field~\cite{Pethick2008}.
The term $\hat H_c$ characterizes the coupling between two hyperfine spin states, where $\Omega$ is the coupling strength.

\begin{figure}[htp]
\begin{center}
\includegraphics[width=1.0\columnwidth]{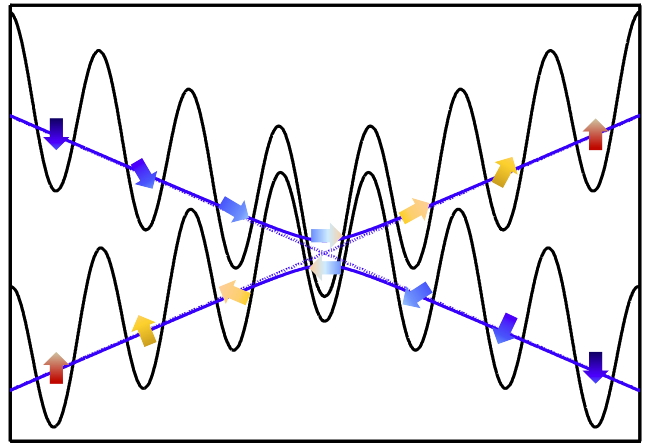}
\end{center}
\caption{(Color online) Schematic diagram of the Wannier-Zeeman system. A spin-up/down atom feels positive/negative Zeeman gradient potential (blue dashed lines). The coupling between two spin states opens the spatial crossing of the two Zeeman energies (blue solid lines), where LZ tunneling may happen when the atom passes the avoided crossing region. The 1D optical lattice potentials are modified by the Zeeman energies and the coupling (black solid lines).}
\label{schematic}
\end{figure}

The Zeeman energy for an atom in $\left|{f, m_f}\right\rangle$ is given as $E_{\textrm{Zeeman}}(z)=-{g_f}{m_f}{\mu _B}Bz$ with the Land$\acute{e}$ factor ${g_f}$, the $z$-component magnetic quantum number $m_f$, and the Bohr magneton ${\mu _B}$.
Because the two spin states have opposite magnetic quantum numbers $m_f$, the two Zeeman energies form a symmetric spatial `scissor-like' structure, see Fig.~\ref{schematic}.
Therefore, the spin-dependent constant force is $\vec{F}_{m_f} = -{{\mathrm d} \over {\mathrm dz}} E_{\textrm{Zeeman}}(z) \vec{e}_z= {g_f}{m_f}{\mu _B}B\vec{e}_z$ with the unit vector $\vec{e}_z$ along $z$-direction and the amplitude $F_{m_f}=|\vec{F}_{m_f}|$.
Due to the coupling between spin states, the spatial `scissor-like' crossing structure is opened and becomes an anti-crossing structure.

\subsection{Single-band tight-binding Hamiltonian}

For a deep lattice potential with a weak spin-dependent force, the tunneling between different bands can be neglected and one can apply single-band and tight-binding approximations to our WZ system (see Appendix A). Thus the system obeys the single-band tight-binding Hamiltonian,
\begin{eqnarray}
\label{equ:single-tight}
  \hat H = && F_{m_f}d {\hat{\sigma} _z} \sum\limits_{j,s}^{} {j\left| {j, s} \right\rangle \left\langle {j,s} \right|} + \varepsilon \sum\limits_{j,s}^{} {\left| {j,s} \right\rangle \left\langle {j,s} \right|} \nonumber \\
  && + \Delta\sum\limits_{j,s}^{} \left( {\left| {j,s} \right\rangle \left\langle {j+1,s} \right|}  + \mathrm{h.c.} \right) \nonumber \\
  && + {{\hbar \Omega } \over 2}\sum\limits_{j}^{} \left({\left| {j,\uparrow} \right\rangle \left\langle {j,\downarrow} \right|} + \mathrm{h.c.} \right).
\end{eqnarray}
Here, $\left|{j,s}\right\rangle$ is the Wannier basis localized in the $j$-th site (in which $s$ indexes the spin state), $d$ is the distance between two neighbour sites, $\Delta$ denotes the hopping strength and $\varepsilon$ the on-site energy (which can be neglected in our calculation).
Obviously, the first term of Hamiltonian (\ref{equ:single-tight}) is the tilts caused by the gradient magnetic field.
For spin states of opposite magnetic moments, their corresponding energy tilts are also opposite.
The last term describes the coupling between two spin states.

Using the Fourier transformation
\begin{equation}\label{equ:Fourier}
\left| {q,s} \right\rangle  = \sqrt {{d \over {2\pi}}} \sum\limits_{j =  -\infty}^{+\infty} {\left| {j,s} \right\rangle {e^{ iqjd}}},
\end{equation}
and the two-component Bloch representation for the quasi-momentum space, as the periodic boundary conditions request $q\Leftrightarrow q+ 2\pi/d$ for the quasi-momentum $q$, one can obtain the Hamiltonian,
\begin{eqnarray}
\label{equ:free}
\hat{H}(q) =
\begin{pmatrix}
{H_{d}^{+}} & {{\hbar\Omega \over 2}} \cr
{{\hbar\Omega \over 2}} & {H_{d}^{-}}
\end{pmatrix},
\end{eqnarray}
with
$$H_{d}^{+} = \left[2\Delta \cos (qd) + i{F_{m_f}}{\partial \over {\partial q}}\right],$$
$$H_{d}^{-} = \left[2\Delta \cos (qd) - i F_{m_f} {\partial \over {\partial q}}\right].$$

\subsection{Energy band structure for the field-free system}

\begin{figure}[!htp]
\begin{center}
\includegraphics[width=1.0\columnwidth]{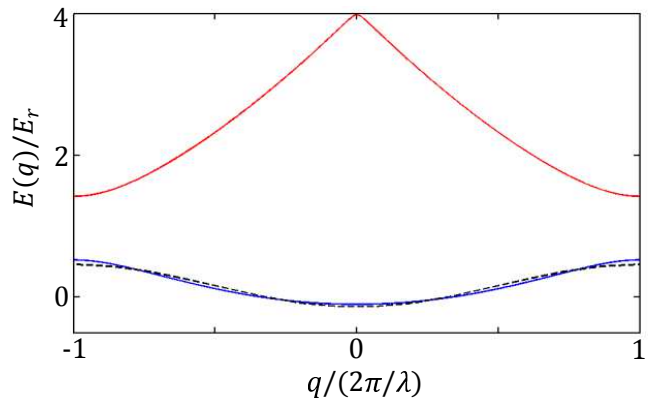}
\end{center}
\caption{(Color online) Band structure for the field-free system. The solid lines are the first two bands for the original Hamiltonian $\hat h_0$ with the lattice depth $V_0 = 1.8 E_r={\hbar}^2 {\kappa}_0^2/(2M)$.
The dash line is the energy band for the corresponding single-band tight-binding Hamiltonian~\eqref{equ:free} with $\Delta= -0.1497 E_r$ and $\mu=0.1635E_r$ calculated from formulae~\eqref{equ:tunneljk}.}
\label{band}
\end{figure}

For the field-free system ($F_{m_f} = 0$ and $\Omega  = 0$),
the Hamiltonian is spin-independent, we have
\begin{equation}
E\left( q  \right) =  2\Delta\cos \left( {q d} \right),
\end{equation}
which indicates that the dispersion relations for spin-up and spin-down atoms are the same.
In Fig.~\ref{band}, we show the dispersion relation of the field-free system (i.e. the spin-independent Hamiltonian $\hat H_0$).
We have compared the band structures of the original Hamiltonian and the corresponding single-band tight-binding Hamiltonian.
It shows that, for the lattice depth $V_0 =1.8E_r={\hbar}^2 {\kappa}_0^2/(2M)$, the first band of the original Hamiltonian almost overlaps with the one of the tight-binding Hamiltonian for the first band.
This means that the tight-binding approximation is valid for such a deep optical lattice potential.

\section{Eigenvalue problem and Wannier-Zeeman states}\label{sec3}

The stationary behavior of a quantum system can be obtained by solving its eigenvalue problem.
For our single-band tight-binding Hamiltonian~(\ref{equ:free}), its $\nu$-th eigenvalue $E_{\nu}$ ($\nu$ is the eigenstate quantum number) and eigenfunctions  $\varphi_{\nu}(q) \equiv \left(\varphi_{\nu,\uparrow}(q),\varphi_{\nu,\downarrow}(q)\right)^T$ ($T$ denotes the transpose of matrix) are given by the coupled first-order differential equations,
\begin{eqnarray}
{ 2\Delta\cos(qd)} {\varphi_{\nu,\uparrow}} + iF_{m_f}{{\mathrm{d} \varphi}_{\nu,\uparrow} \over \mathrm{d}q} + {{\hbar\Omega} \over 2} {\varphi_{\nu,\downarrow}} = E_{\nu}{\varphi_{\nu,\uparrow}},\label{eigeneq01} \\
{ 2\Delta\cos(qd)} {\varphi_{\nu,\downarrow}} - iF_{m_f}{{\mathrm{d} \varphi}_{\nu,\downarrow} \over \mathrm{d}q} +{{\hbar\Omega} \over 2} {\varphi_{\nu,\uparrow}} = E_{\nu}{\varphi_{\nu,\downarrow}}. \label{eigeneq02}
\end{eqnarray}
Obviously, the eigenfunctions are periodic functions satisfying $\varphi_{\nu,s}(q)=\varphi_{\nu,s}(q+2\pi/d)$. To simplify Eqs.~\eqref{eigeneq01} and \eqref{eigeneq02}, we introduce the following transformation,
\begin{eqnarray}
&&  {\varphi _ {\nu,\uparrow }}(q) = {e^{+i{{{d_0}} \over d}\sin qd}}\tilde{\varphi} _ {\nu,\uparrow }(q), \label{eigenstate1}  \\
&&  {\varphi _ {\nu,\downarrow }}(q) = {e^{-i{{{d_0}} \over d}\sin qd}\tilde{\varphi} _ {\nu,\downarrow }(q)}, \label{eigenstate2}
\end{eqnarray}
with $d_0=2\Delta/F_{m_f}<0$. Substitute Eqs.~\eqref{eigenstate1}-\eqref{eigenstate2} into Eqs.~\eqref{eigeneq01}-\eqref{eigeneq02}, we obtain
\begin{eqnarray}
 +iF_{m_f}{\mathrm{d} \over {\mathrm{d}q}}\tilde{\varphi} _ {\nu,\uparrow }+{\hbar \Omega \over {2}}e^{-i2{d_0\over {d}}\sin{qd}}\tilde{\varphi} _ {\nu,\downarrow }=E_{\nu} \tilde{\varphi} _ {\nu,\uparrow }, \label{eigenstate3} \\
 -iF_{m_f}{\mathrm{d} \over {\mathrm{d}q}}\tilde{\varphi} _ {\nu,\downarrow }+{\hbar \Omega \over {2}}e^{+i2{d_0\over {d}}\sin{qd}}\tilde{\varphi} _ {\nu,\uparrow}=E_{\nu} \tilde{\varphi} _ {\nu,\downarrow}. \label{eigenstate4}
\end{eqnarray}
The trial solutions of $\tilde{\varphi}_{\nu,\uparrow}(q)$ and $\tilde{\varphi}_{\nu,\downarrow}(q)$  can be expanded as Fourier series
\begin{eqnarray}
&&\tilde{\varphi}_{\nu,\uparrow}=\sum\limits_{m=-M}^{+M} {{A_{\nu,m}}{e^{ - iqmd}}},\label{tial1} \\
&&\tilde{\varphi}_{\nu,\uparrow}=\sum\limits_{m=-M}^{+M} {{B_{\nu,m}}{e^{iqmd}}},\label{tial2}
\end{eqnarray}
where $M$ is the truncation of the Fourier series. Substitute Eqs.~\eqref{tial1}-\eqref{tial2} into Eqs.~\eqref{eigenstate3}-\eqref{eigenstate4}, one can find that the coefficients $A_{\nu,m}$ and $B_{\nu,m}$ obey the recurrence relations,
\begin{eqnarray}
\sum\limits_m^{} {{{\hbar \Omega } \over 2}{{( - 1)}^{m + m'}}{J_{m + m'}}( - 2{{{d_0}} \over d}){B_{\nu,m}}} && \nonumber\\
+ m'F_{m_f}d{A_{\nu,m'}} = &E_{\nu}{A_{\nu,m'}},& \label{coeffient11}
\end{eqnarray}
\begin{eqnarray}
\sum\limits_m^{} {{{\hbar \Omega } \over 2}{{( - 1)}^{m + m'}}{J_{m + m'}}( - 2{{{d_0}} \over d}){A_{\nu,m}}} && \nonumber\\
+ m'F_{m_f}d{B_{\nu,m'}} = &E_{\nu}{B_{\nu,m'}},& \label{coeffient12}
\end{eqnarray}
where $J_{m + m'}( - 2{{{d_0}} \over d})$ are the Bessel functions of the first kind. The eigenvalues $E_{\nu}$ and the coefficients $A_{\nu,m}$, $B_{\nu,m}$ can be obtained by solving the Eqs.~\eqref{coeffient11}-\eqref{coeffient12}.

Using the Fourier transformation, one can find the eigenstates in the Wannier representation,
\begin{eqnarray}
{\phi_{\nu, \uparrow}}(j) &=& \int_{-{\pi \over d}}^{\pi \over d} {\mathrm{d}q\left\langle {{j,\uparrow}} \mathrel{\left|{\vphantom {{j,\uparrow} {q,\uparrow}}}\right. \kern-\nulldelimiterspace} {{q,\uparrow}} \right\rangle \left\langle {{q,\uparrow}} \mathrel{\left|{\vphantom{{q,\uparrow} {{\varphi_{\nu}}}}} \right. \kern-\nulldelimiterspace} {{{\varphi_{\nu}}}} \right\rangle } \nonumber\\
&=& \sum\limits_{m}^{}{{{A_{\nu,m} d} \over {2\pi }}\int_{ - {\pi \over d}}^{\pi \over d} {\mathrm{d}q ~\textrm{e}^{i \left[ {q(j - m)d + {{{d_0}} \over d}\sin (qd)} \right]}}} \nonumber \\
&=& \sum\limits_m^{}{A_{\nu,m} {J_{j - m}}\left( { - {{{d_0}} \over d}} \right)},
\end{eqnarray}
and
\begin{equation}
 {\phi_{\nu, \downarrow}}(j) = \sum \limits_m^{} {B_{\nu,m} {J_{j + m}}\left( { {{{d_0}} \over d}} \right)}.
\end{equation}
Here $J_{j-m}(-d_0/d),J_{j+m}(d_0/d)$ are the Bessel functions.
If $\left|d_0/d\right|$ is sufficiently small, the Bessel functions $J_{j-m}(-d_0/d)$ and $J_{j+m}(d_0/d)$ well localize in the lattice sites of $m + d/{d_0} < j < m - d/{d_0}$ and  $- m + d/{d_0} < j <  - m - d/{d_0}$, respectively.

\begin{figure}[!htp]
\begin{center}
\includegraphics[width=1.0\columnwidth]{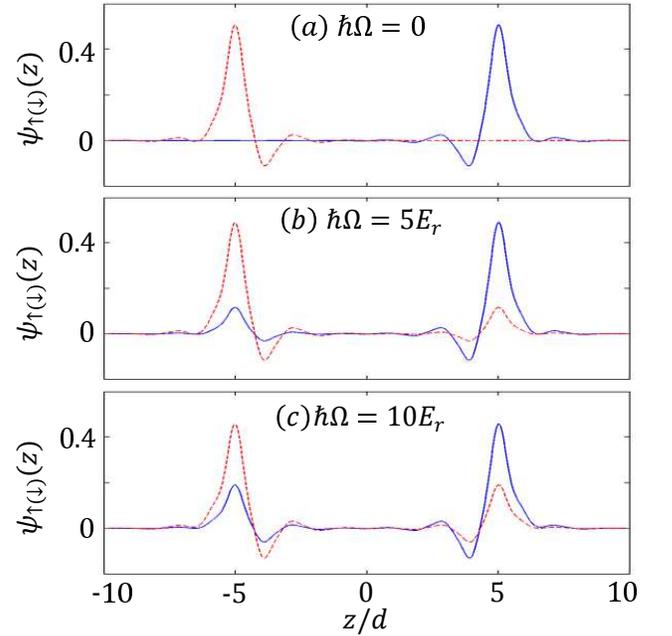}
\end{center}
\caption{(Color online) Wannier-Zeeman states for different inter-spin coupling strengths: (a) $\hbar\Omega=0$, (b) $\hbar\Omega=5E_r$ and (c) $\hbar\Omega=10E_r$. The blue solid lines and red dashed lines correspond to the spin-up and spin-down components, respectively. The parameters are chosen as $M=50$, $\Delta=-0.15E_r$, $Fd=E_r$.}
\label{eigenstate}
\end{figure}

Thus, the eigenstates for the system in the two-component Wannier basis can be written as,
\begin{equation}
  \left| {{\psi _{\nu}}} \right\rangle  = { \begin{pmatrix}
\sum\limits_{j}^{} {{\phi _{\nu, \uparrow }}(j) }  \left|j,\uparrow\right\rangle \cr
\sum\limits_{j}^{} {{\phi _{\nu, \downarrow }}(j) \left|j,\downarrow\right\rangle }
\end{pmatrix}}.
  \label{WZS_spinup_spindown1}
\end{equation}
In the coordinate space, the eigenfunctions is given by
\begin{equation}
\psi _{\nu}(z)
=  \left\langle {z}
 \mathrel{\left | {\vphantom {z {{\psi _{\nu}}}}}
 \right. \kern-\nulldelimiterspace}
 {{{\psi _{\nu}}}} \right\rangle
= \begin{pmatrix}
   {{\psi _{\nu, \uparrow }}(z)}  \cr
   {{\psi _{\nu, \downarrow }}(z)}
  \end{pmatrix}
= \begin{pmatrix}
{\sum\limits_j^{} {{\phi _{\nu, \uparrow }}(j){w_{j}}(z)} }  \cr
{\sum\limits_j^{} {{\phi _{\nu, \downarrow }}(j){w_{j}}(z)} }
\end{pmatrix},\nonumber
\end{equation}
where ${w_{j}}(z) = {w_{j,s}}(z)=\left\langle {z}
 \mathrel{\left | {\vphantom {z {j,s}}}
 \right. \kern-\nulldelimiterspace}
 {{j,s}} \right\rangle  $ is the $j$-th Wannier functions of the lowest band. To explore how the inter-spin coupling strength $\Omega$ affects eigenfunctions, we calculate eigenfunctions for different values of $\Omega$ by using the numerical methods in Ref.~\cite{Walters2013,Arash2008}.
In Fig.~3,  we show the 112-th eigenstate in the coordinate space for different values of $\Omega$.
The Fourier series are truncated at $M=50$ and the parameters are chosen as $\Delta=-0.15E_r$, $Fd=E_r$, and $\hbar\Omega=(0, 5E_r, 10E_r)$.
The eigenstates inherit the symmetry of the Hamiltonian. If the inter-spin coupling is absent (i.e. $\hbar\Omega=0$), the spin-up component localizes at $l=+5$ while the spin-down component symmetrically localizes at $l=-5$, see Fig.~\ref{eigenstate} (a).
Due to the inter-spin coupling, the spin-up and spin-down components are mixed with each other and the mixing becomes more significant when $\Omega$ increases, see Fig.~\ref{eigenstate}~(b) for $\hbar\Omega=5E_r$ and Fig.~\ref{eigenstate}~(c) for $\hbar\Omega=10E_r$.

For the case of $\Omega=0$, the spin-up component and the spin-down component are decoupled and so that the eigen-problem can be described by the following two independent equations,
\begin{eqnarray}
{ 2\Delta\cos(qd)} {\varphi_{\uparrow}} + iF_{m_f}{{\mathrm{d} \varphi}_{\uparrow} \over \mathrm{d}q} = E{\varphi_{\uparrow}},\\
{ 2\Delta\cos(qd)} {\varphi_{\downarrow}} - iF_{m_f}{{\mathrm{d} \varphi}_{\downarrow} \over \mathrm{d}q} = E{\varphi_{\downarrow}}.
\end{eqnarray}
The eigenvalues for the above two equations form two independent WS ladders,
\begin{eqnarray}\label{WZL}
 E_{n,\uparrow} &=& n F_{m_f}d,~~~~~~(n=0,\pm 1,\pm 2,...),\\
 E_{n',\downarrow} &=& n' F_{m_f}d,~~~~~~(n'=0,\pm 1,\pm 2,...),
\end{eqnarray}
which respectively correspond to two different series of WS states for the spin-up and spin-down components \cite{Hartmann2004,Holthaus1996},
\begin{eqnarray}
&{\varphi_{n,\uparrow}(q) } = \sqrt {{d \over {2\pi}}} \exp\left[{-iqnd +i{{{d_0}} \over d} \sin(qd)}\right]
\begin{pmatrix}
{1} \cr
{0}
\end{pmatrix}
,\label{WZS1}  \\
&{\varphi_{n',\downarrow}(q) } =  \sqrt {{d \over {2\pi}}} \exp\left[{+iqn'd - i{{{d_0}} \over d} \sin(qd)}\right]
\begin{pmatrix}
{0}  \cr
{1}
\end{pmatrix}\label{WZS2}.
\end{eqnarray}
The stable states, which have time-independent probability distributions, can be arbitrary superpositions of the WS states~(\ref{WZS1}) and (\ref{WZS2}),
\begin{equation}
  \varphi_{n,n'}=A \varphi_{n,\uparrow}(q) + B \varphi_{n',\downarrow}(q),
\end{equation}
which read as
\begin{eqnarray}
\left| {{\psi _{n,n'}}} \right\rangle
=  \begin{pmatrix}
{A \sum\limits_{j}^{} {{J_{j - n}}\left({-{{{d_0}} \over d}}\right) \left|j, \uparrow\right\rangle}}  \cr
{B \sum\limits_{j}^{} {{J_{j + n'}}\left({+{{{d_0}} \over d}}\right) \left|j, \downarrow\right\rangle}}
\end{pmatrix},\nonumber
  \label{WZS_spinup_spindown2}
\end{eqnarray}
in the two-component Wannier representation.

Here, the complex coefficients $A$ and $B$ satisfy the normalization condition $\left|A\right|^2+\left|B\right|^2=1$.
When the lattices are sufficiently deep or gradient magnetic field is sufficiently strong, the spin-up component is localized at $n$-th site and the spin-down component is localized at $(-n')$-th site.
Due to absence of inter-spin coupling, both spatial distributions and spin degrees of freedom of these stable states will not change with time.
The energy expectations for these stable states are given as $\bar{E}=\left|A\right|^2E_{n,\uparrow}+\left|B\right|^2 E_{n',\downarrow}$ and the corresponding uncertainties read as $\Delta E = \sqrt{\langle\hat H^2\rangle-\langle\hat H\rangle ^2}=\left|AB(E_{n,\uparrow}-E_{n',\downarrow})\right|$.
If and only if $A=0$ or $B=0$ or $n=n'$, the uncertainty $\Delta E=0$ and the corresponding stable state becomes an eigenstate for the uncoupled system.

\section{Dynamics of Bloch oscillations and Landau-Zener tunneling}\label{sec4}

In this section, we consider the dynamics of a spin-half particle passing the spatial anti-crossing formed by the spin-independent tilted optical lattices and the coupling between two spin states.
In particular, we will discuss how the inter-spin coupling affects Bloch oscillations and how Bloch oscillations induce inter-spin LZ tunneling.
Different from the inter-band LZ tunneling in a WS system of a spinless particle, which is induced by the tilt, the inter-spin LZ tunneling in our WZ system is caused by the joint effects of the inter-spin coupling and the Bloch oscillations.

An arbitrary state for Hamiltonian~\eqref{equ:single-tight} can be expanded as a linear superposition of Wannier states for different lattice sites,
\begin{equation}
\left| {\varphi \left( t \right)} \right\rangle
= {\begin{pmatrix}
\sum_{j = -N}^{+N} {{c_{j, \uparrow }}(j) }  \left|j,\uparrow\right\rangle \cr
\sum_{j = -N}^{+N} {{c_{j, \downarrow }}(j) \left|j,\downarrow\right\rangle }
\end{pmatrix}}.
\end{equation}
If the hopping strength $\Delta$ is very small compared to the Rabi frequency $\Omega$ and the nearest neighbouring energy jump $F_{m_f}d$, one can approximately decouple the correlation between different lattice sites by ignoring the hopping terms.
Thus, the subsystem for the $j$-th lattice site can be described by the following Hamiltonian matrix,
\begin{equation}
{\hat{H}}_j=
\begin{pmatrix}
{+{F_{m_f}}dj} & {{\hbar \Omega  \over 2}} \cr
{{\hbar \Omega  \over 2}} & {-{F_{m_f}}dj}
\end{pmatrix},
\label{Onsite_Ham}
\end{equation}
with two eigenvalues,
\begin{equation}
  {E_{j, \pm }} =  \pm {\Gamma_j},
\end{equation}
for two dressed states,
\begin{equation}
  {\psi _{ j, +}} =
  \begin{pmatrix}
   {{{\sqrt {{\Gamma_j} + {F_{m_f}}dj} } \over {\sqrt {2{\Gamma_j}} }}}  \cr
   {{{\sqrt {{\Gamma_j} - {F_{m_f}}dj} } \over {\sqrt {2{\Gamma_j}} }}}
 \end{pmatrix},\;
 {\psi _{ j, - }}=
 \begin{pmatrix}
   {{{ - \sqrt {{\Gamma_j} - {F_{m_f}}dj} } \over {\sqrt {2{\Gamma_j}} }}} \cr
   {{{\sqrt {{\Gamma_j} + {F_{m_f}}dj} } \over {\sqrt {2{\Gamma_j}} }}}
 \end{pmatrix}.
 \label{equ:dress-state}
\end{equation}
Here ${\Gamma _j} = \sqrt {{{\left( {{F_{m_f}}dj} \right)}^2} + {{\left( {{{\hbar \Omega } / 2}} \right)}^2}}$.
The eigenvalue set $\left\{E_{j, +}, E_{j, -}\right\}$ form two spatial energy ladders: the upper $V$-shape ladder and the lower $\Lambda$-shape ladder.
The WZ ladder~\eqref{WZL} corresponds to asymptotes of these two ladders.
In the region far away from the spatial anti-crossing, where $\left| {{F_{m_f}}dj} \right| \gg {{\hbar \Omega } \over 2}$, the eigenvalues can be approximately given as the WZ ladders~\eqref{WZL}.
Correspondingly, the dressed states can be approximately written as
\begin{equation}
\begin{split}
 & {\psi_{j, +}} \approx {\varphi_{j, \uparrow}},
 ~~{\psi_{j, -}} \approx {\varphi_{j, \downarrow}}
 \;\;{\rm{if}}\;\;j \rightarrow +\infty,\\
 & {\psi_{j, +}} \approx {\varphi_{j, \downarrow}},
 ~~{\psi_{j, -}} \approx  - {\varphi_{j, \uparrow}}
 \;\;{\rm{if}}\;\;j \rightarrow -\infty.
\end{split}
\end{equation}
Naturally, in addition to the bare Wannier basis $\left\{ {\left| {j,\uparrow} \right\rangle, \left| {j, \downarrow} \right\rangle} \right\}$, one may use the dressed Wannier basis $\left\{ {\left| {j, +} \right\rangle, \left| {j, -} \right\rangle} \right\}$ to analyze the dynamics of the WZ system.

In the bare Wannier basis, the dynamics obeys the time-dependent Schr\"{o}dinger equation
\begin{equation}
  i\hbar {\partial \over {\partial t}} \varphi(t) = \hat H \varphi(t),
\end{equation}
with $\hat H$ given by Eq.~\eqref{equ:single-tight}.
For simplicity, we assume that the system is prepared in spin-up state with a Gaussian spatial distribution, that is,
 \begin{equation}
\label{oscillations}
\begin{cases}
{c_{j, \uparrow}}\left( 0 \right) = f\exp \left( { - {{{{\left( {j - {j_0}} \right)}^2}} \over {2{\sigma ^2}}} + i{q _0}jd} \right), \\
{c_{j, \downarrow}}\left( 0 \right) = 0.
\end{cases}
\end{equation}
Here $f$ is a normalization factor, ${j_0}$ is the wave-packet center, $\sigma$ is the standard deviation of Gaussian wave-packet, ${q_0}$ denotes the initial quasi-momentum.
Below, for different coupling strengths $\Omega$, we show the time-evolution dynamics of the initial state \eqref{oscillations}. In our calculations, the other parameters are chosen as $\Delta =-0.15{E_r}$, $F_{m_f}d = 0.01{E_r}$, $j_0=45$, $\sigma = 5$ and ${q _0}=0$.

\subsection{Bloch oscillations}

\begin{figure}[!htp]
\begin{center}
\includegraphics[width=1.0\columnwidth]{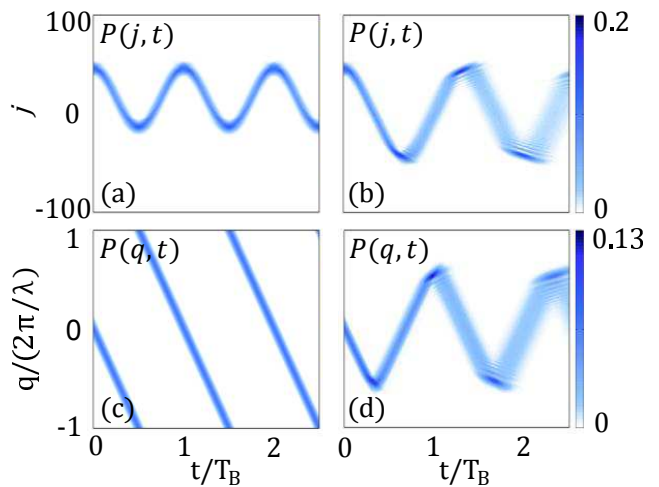}
\end{center}
\caption{(Color online) Bloch oscillations (left column) for $\hbar\Omega=0$ and adiabatic BLZ dynamics (right column) for $\hbar\Omega=0.2E_r$. The total spatial probability distributions $P(j,t)$ for the Bloch oscillations and the adiabatic BLZ dynamics are shown in (a) and (b), respectively. The corresponding total quasi-momentum probability distributions $P(q,t)$ are shown in (c) and (d). The other parameters are chosen as $\Delta = -0.15{E_r}$, $F_{m_f}d = 0.01{E_r}$ and $j_0=45$.}
\label{adiabatic-dynamics}
\end{figure}

If the inter-spin coupling is absent or sufficiently weak, the coupling effects can be neglected.
Similar to the case of spinless particles~\cite{Hartmann2004}, the dynamics in the uncoupled WZ system can be analytically described.
In the first Brillouin zone $\left|q\right| \le \pi/d$, the quasi-momentum $q$ obeys the following acceleration relation
\begin{equation}\label{fequalma}
q(t) =  q_0 \mp {F_{m_f}t \over \hbar}.
\end{equation}
Here, the symbols ``$-$" and ``$+$" correspond to spin-up and spin-down components, respectively.
Because the spin-up atom and the spin-down atom feel opposite constant forces, they are accelerated toward opposite directions.
Once a wave-packet reaches the boundary of the first Brillouin zone, it will reenter the first Brillouin zone from the other side.
Consequently, the group velocity ${v_g} = {1 \over \hbar }{{\partial E\left( q  \right)} \over {\partial q }}$ changes its direction at the boundary and the wave-packet undergoes Bloch oscillations of period ${T_B} = 2\pi \hbar /\left( {F_{m_f}d} \right)$.
In Fig.~\ref{adiabatic-dynamics}~(a) and (c), we show the Bloch oscillations in the uncoupled WZ system.
Due to the tilt induced by the magnetic field gradient, periodic oscillations appear in the total probability distribution,
\begin{equation}\label{spd}
P(j,t) = P_{\uparrow}(j,t)+P_{\downarrow}(j,t).
\end{equation}
Here $P_{\alpha}(j,t)={\left| {\left\langle{j, \alpha}|{\varphi (t)}\right\rangle} \right|^2}$ is the probability of finding the particle in the Wannier state $\left| j, \alpha \right\rangle$ at time $t$.
By using the Fourier transformation \eqref{equ:Fourier}, one can obtain the total probability distribution in the quasi-momentum space,
\begin{equation}\label{ppd}
P(q,t) = P_{\uparrow}(q,t)+P_{\downarrow}(q,t).
\end{equation}
with $P_{\alpha}(q,t)={\left| {\left\langle{q, \alpha}|{\varphi (t)}\right\rangle} \right|^2}$ denoting the probability of finding the particle in $\left|q, \alpha \right\rangle$ at time $t$.
Because there is no coupling between the two spin states, the dynamics of the spin-up component and the one of the spin-down component are independent and so that the uncoupled WZ system can be understood as two independent WS systems with opposite tilts.

Furthermore, we find that perfect Bloch oscillations may also appear in the region far from the spatial anti-crossing in the WZ system of significant inter-spin coupling.
This is because that the large energy gap between two spin states in that region inhibits the spin population transfer even there is significant inter-spin coupling.

\subsection{Bloch-Landau-Zener dynamics}

In the WZ system with inter-spin coupling, the spatial crossing between the two WS systems for the two spin components becomes an anti-crossing.
The spin dynamics of a particle across the spatial anti-crossing depends on the gap of the spatial anti-crossing and the inter-spin LZ tunneling may take place.
In particular, Bloch oscillations and inter-spin LZ tunneling may mix together. We name such a type of dynamics of mixed Bloch oscillations and inter-spin LZ tunneling as Bloch-Landau-Zener (BLZ) dynamics.
Different from the Bloch-Zener transitions between two different energy bands around the boundary of Brillouin zones~\cite{Breid2006, Dreisow2009}, our BLZ dynamics involves only a single energy band and happens between local dressed states around the spatial anti-crossing region.
Here, we will study the BLZ dynamics and derive an analytical formula for the inter-spin LZ transition in our BLZ dynamics.

In our WZ system, the spatial anti-crossing region is around the center ($l=0$), where inter-spin LZ transitions may happen.
Therefore, the BLZ dynamics appear if the Bloch oscillations pass through the center.
Due to the two spin components have opposite group velocities, the BLZ dynamics may appear if a fully spin-polarized particle is initially situated in the right/left region not far away from the center.
Without loss of generality, as an example, we consider a spin-up particle in the right region ($j_0>0$) and the initial velocity is zero ($q_0=0$).
The group velocity of the wave-packet is given by
\begin{equation}
  {v_g} = {1 \over \hbar }{{\partial E(q)} \over {\partial q}} = {{2\Delta d} \over \hbar }\sin \left( {{{{F_{m_f}}d} \over \hbar }t} \right).
\end{equation}
By integrating the group velocity over time, the position of the wave-packet is given as
\begin{equation}
\begin{split}
  jd &= j_0 d + \int_0^{{t}} {{v_g}(\tau ) \mathrm{d} \tau}   \\
  &= j_0 d - {{2\Delta } \over {{F_{m_f}}}}\left[ {\cos \left( {{{{F_{m_f}}d} \over \hbar }{t}} \right) - 1} \right].
\end{split}
\label{position}
\end{equation}
The appearance of BLZ dynamics requires $jd \le 0$ at some time $t$, thus the occurrence condition for the BLZ dynamics reads as
\begin{equation}
  j_0 d \le D_{pi}=-{{4 \Delta} \over {F_{m_f}}}.
\end{equation}
Here, $D_{pi}=-{{4 \Delta} \over {F_{m_f}}}$ is the populating interval of Bloch oscillations~\cite{Hartmann2004}.
If the initial distance from the wave-packet to the center is larger than $D_{pi}$, the wave-packet will mainly undergo Bloch oscillations and there is no significant inter-spin LZ transitions.
Otherwise, if the initial distance is smaller than $D_{pi}$, significant inter-spin LZ transitions take place when the particle goes through the anti-crossing region.

According to Eq.~\eqref{position}, given $jd=0$, one can obtain the occurrence times for the first inter-spin LZ transitions,
\begin{equation}
  {t_1} = {{{T_B}} \over {2\pi }}\arccos \left( {{{{j_0}{F_{m_f}}d} \over {2\Delta }} + 1} \right).   \label{equ:t1}
\end{equation}
Making a Taylor expansion of $jd$ around the time $t_1$ and keeping the linear term, one has
\begin{equation}
  jd ={{2\Delta d} \over \hbar }\sin \left( {{{{F_{m_f}}d} \over \hbar }{t_1}} \right)t'+\mathcal{O}(t'^2)= v_1t',
  \label{taylor}
\end{equation}
where $v_1={{2\Delta d} \over \hbar }\sin \left( {{{{F_{m_f}}d} \over \hbar}{t_1}} \right)$ and $t'=t-t_1$.
Substituting Eq.~\eqref{taylor} into Eq.~\eqref{Onsite_Ham}, one gets an instantaneous Hamiltonian for the first inter-spin LZ transition,
\begin{equation}
{\hat{H}}(t')=
\begin{pmatrix}
{+{F_{m_f}}v_1t'} & {{\hbar \Omega  \over 2}} \cr
{{\hbar \Omega  \over 2}} & {-{F_{m_f}}v_1t'}
\end{pmatrix}.
\end{equation}
Applying the conventional LZ formula for a two-level system~\cite{Landau1981Quantum}, the transition probability from dressed-`+' states to dressed-`-' states in the first LZ transition is given as
\begin{equation}
{P_{1,+\rightarrow-}} = \exp \left( { - \pi {{{(\hbar\Omega) ^2}} \over {4 \hbar \left|F_{m_f} v_1\right|}}} \right).
  \label{LZ+-}
\end{equation}
It clearly shows that the transition probability $P_{1,+\rightarrow-}$ tends to $0$ when the inter-spin coupling strength $\hbar \Omega \rightarrow \infty$.
Similarly, the transition probability for the second LZ transition can be given analytically.

\subsubsection{Adiabatic dynamics}

In the strong coupling regime, the gap between the spatial anti-crossing is large and the internal spin dynamics undergoes adiabatic evolution along the dressed states. Due to the large gap, the local energy levels can be approximately given as the WZ ladders~\eqref{WZL}.
We consider an initial state~\eqref{oscillations} center around $j_0=45$, which satisfies the condition $0<j_{0}d<D_{pi}$ for the BLZ dynamics.
In Fig.~\ref{adiabatic-dynamics}, we show the time-evolution of the total probability distribution for the Bloch oscillations and the BLZ dynamics.
Surprisingly, the total probability distribution of the WZ system still undergo oscillations similar to Bloch oscillations.

\begin{figure}[!htp]
\begin{center}
\includegraphics[width=1.0\columnwidth]{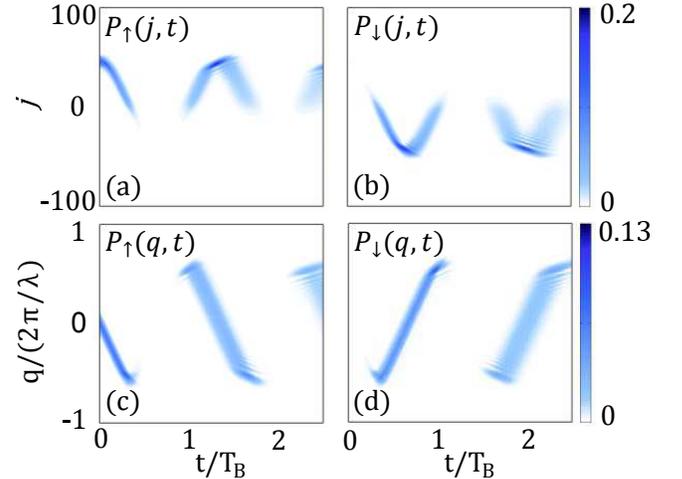}
\end{center}
\caption{(Color online) Spin dynamics under strong inter-spin coupling. Top: The spatial dynamics of the spin-up component (a) and the spin-down component (b). Bottom: The quasi-momentum dynamics of the spin-up component (c) and the spin-down component (d). The parameters are chosen as the same ones for Fig.~\ref{adiabatic-dynamics}~(b).}
\label{adiabatic-up-down}
\end{figure}

To understand how inter-spin population transfer occurs in our WZ system, we calculate the time-evolution of two spin components. In Fig.~\ref{adiabatic-up-down}, we show the time-evolution of $P_{\alpha}(j,t)={\left| {\left\langle{j, \alpha}|{\varphi (t)}\right\rangle} \right|^2}$ and $P_{\alpha}(q,t)={\left| {\left\langle{q, \alpha}|{\varphi (t)}\right\rangle} \right|^2}$.
The time-evolution includes three types of dynamical processes: (i) Bloch oscillation of the spin-up component in the right region ($0<jd<D_{pi}$) away from the anti-crossing region around $j=0$, (ii) spin inversion during the anti-crossing region, and (iii) Bloch oscillation of the spin-down component in the left region ($-D_{pi}<jd<0$) away from the anti-crossing region.

As the initial state is a spin-up wave-packet in the right region away from the anti-crossing region, according to Eqs.~\eqref{oscillations} and \eqref{fequalma}, the dynamics is firstly dominated by the Bloch oscillation of the spin-up component and the wave-packet moves toward the anti-crossing region.
In the anti-crossing region, due to the strong inter-spin coupling, spin inversion takes place, that is, the spin degree of freedom changes from $\left|\uparrow\right\rangle$ to $\left|\downarrow\right\rangle$.
As the spin inversion process does not alter the wave-packet velocity, the wave-packet continuously moves toward the left region after the spin population inversion.
In the left region away from the anti-crossing region, the dynamics is dominated by the Bloch oscillation of the spin-down component and the quasi-momentum continuously varies from the negative maximum to the positive maximum.
After the spin-down component changes its group velocity direction, the wave-packet moves back toward the anti-crossing region and the spin inversion takes place again in the anti-crossing region.
Then the wave-packet moves in the right region and the dynamics is again dominated by the Bloch oscillation of the spin-up component and the quasi-momentum continuously varies from the positive maximum to the negative maximum.
Repeating the above processes again and again, the system undergoes oscillations similar to Bloch oscillations, see Fig.~\ref{adiabatic-dynamics}~(a) and (b).

However, the oscillations under strong inter-spin coupling are very different from the ones in a WS system or a uncoupled WZ system.
In a WS system or a uncoupled WZ system, the acceleration keeps its direction unchanged and so that the quasi-momentum jumps from the negative maximum to the positive maximum (or vice versa), see Fig.~\ref{adiabatic-dynamics}~(c).
Under strong inter-spin coupling, the acceleration changes its direction in the spin inversion process.
Thus the quasi-momentum may have no jumps if the Brillouin zone boundary is not crossed before the acceleration changes its direction, see Fig.~\ref{adiabatic-dynamics}~(d) and Fig.~\ref{adiabatic-up-down}.
In addition, the oscillation period may be different from the Bloch period for the uncoupled WZ system.
If the oscillation does not pass through the anti-crossing region, the oscillation period is just the Bloch period.
Otherwise, if the oscillation passes through the anti-crossing region, due to the symmetry of our coupled WZ system, the oscillation period can be approximately estimated as four times of $t_1$ given by Eq.~\eqref{equ:t1}.

\begin{figure}[!htp]
\begin{center}
\includegraphics[width=1.0\columnwidth]{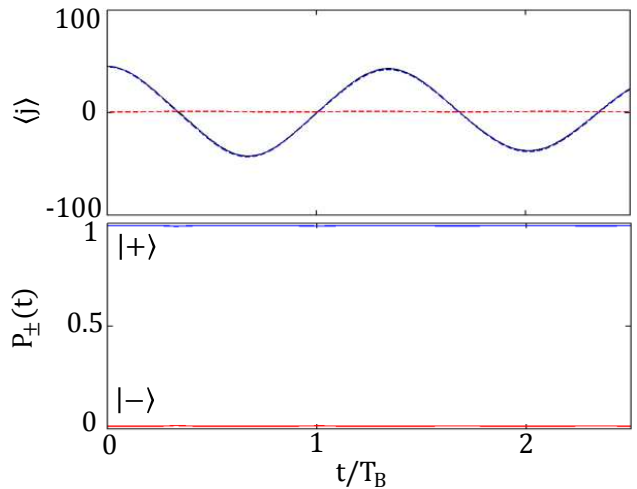}
\end{center}
\caption{(Color online) Time-evolution of the mean positions and the dressed-state populations.
Top: the mean positions of the whole wave-packet and the mean positions of two dressed components.
Bottom: the adiabatic dynamics of dressed-state populations.
The parameters are chosen as the same ones for Fig.~\ref{adiabatic-dynamics}~(b).}
\label{adiabatic-dressed-state}
\end{figure}

To explore whether the spin dynamics is adiabatic, we analyze the population dynamics of the local dressed states~(\ref{equ:dress-state}).
Similar to the conventional LZ problem of a spin-half particle, the adiabaticity of our WZ system may be examined by whether the time-evolution follows the local eigenstates (i.e. the local dressed states).
That is to say, the WZ system undergoes adiabatic spin evolution if its spin degrees of freedom follow the local dressed states~(\ref{equ:dress-state}).
As the passage through the spatial anti-crossing is driven by Bloch oscillations, we call the dynamics in such an adiabatic LZ process as adiabatic BLZ dynamics.

We have numerically calculated the mean position of the whole wave-packet,
\begin{equation}
\left\langle j \right\rangle = \sum\limits_j^{} {{P}(j,t)j},
\end{equation}
and the mean positions of two dressed components,
\begin{equation}
  \left\langle {{j_\pm }} \right\rangle = \sum\limits_j^{} {{P_{j, \pm}}(t)j}.
\end{equation}
Here $P(j,t) =P_{j, \uparrow}(t)+P_{j, \downarrow}(t) =P_{j,+}(t)+P_{j,-}(t)$ and $P_{j, \pm}(t)={\left| {\langle{j, \pm}} |{\varphi(t)}\rangle \right|^2}$.
Our numerical results show the mean position of the whole wave-packet (black solid line) almost fully overlaps with the mean position of the dressed-`+' component (blue dashed line), see Fig.~\ref{adiabatic-dressed-state}.
Furthermore, we also calculate the populations in the two dressed components $P_{\pm}(t)=\sum_j {{P_{ j, \pm}}(t)}$ and find that they keep as their initial values unchanged.

According to the analytical formula~\eqref{LZ+-}, as the parameters $\hbar\Omega=0.2~E_r$, $F_{m_f}d = 0.01~{E_r}$ and $\hbar v_1=-0.0026~d E_r$, the transition probability for the first inter-spin LZ transition is approximately equal to $5.604 \times 10^{-6}$, which is well consistent with the numerical estimation.
Such a small transition probability means that the internal spin degrees of freedom adiabatically follow the local dressed states.

\subsubsection{Non-Adiabatic dynamics}

If the inter-spin coupling is not sufficiently strong, the gap between the spatial anti-crossing is not large enough for accomplishing complete spin inversion.
In the context of the dressed state basis, the dressed-state populations $P_{\pm}(t)$ change between each other when the system goes through the spatial anti-crossing.
That is to say, by using the non-adiabatic LZ tunneling, the spatial anti-crossing plays the role of a beam splitter dependent on the inter-spin coupling strength.

\begin{figure}[!htp]
\begin{center}
\includegraphics[width=1.0\columnwidth]{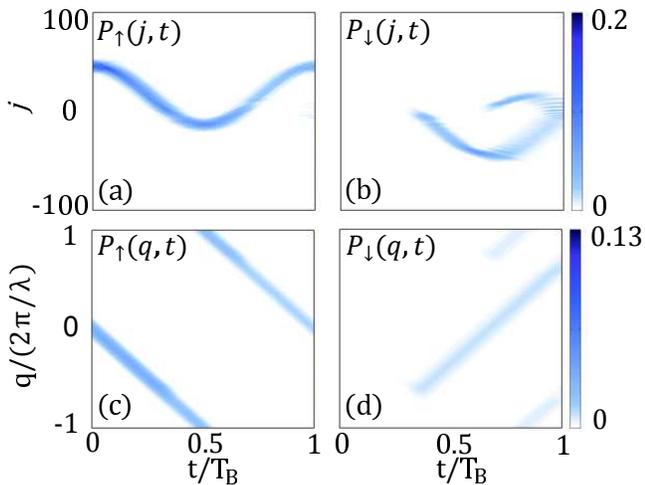}
\end{center}
\caption{(Color online) Spin dynamics under non-strong inter-spin coupling. Top: The spatial dynamics of the spin-up component (a) and the spin-down component (b). Bottom: The quasi-momentum dynamics of the spin-up component (c) and the spin-down component (d).
The parameters are chosen as $\Delta = -0.15{E_r}$, $F_{m_f}d = 0.01{E_r}$, $\hbar \Omega = 0.04{E_r}$ and $j_0=45$. }
\label{non-adiabatic-up-down}
\end{figure}

To understand the spin dynamics under non-strong inter-spin coupling, we simulate the time evolution of a spin-up wave-packet center around $j_0=45$ and analyze the dynamics of spin distribution, see Fig.~\ref{non-adiabatic-up-down}. Obviously, such an initial wave-packet satisfies the condition $0<j_{0}d<D_{pi}$ for the BLZ dynamics.
In the right region ($0<jd<D_{pi}$), the wave-packet firstly undergoes Bloch oscillation towards the anti-crossing region.
In the anti-crossing region, the spin-up component partially changes into the spin-down component.
Away from the anti-crossing region, due to the large gap, the spin-up wave-packet and the spin-down one almost undergo independent Bloch oscillations.
According to Eq.~\eqref{fequalma}, the two spin components have opposite acceleration directions. This is confirmed by the time evolution in quasi-momentum space, the spin-up component moves toward the negative direction while the spin-down component moves toward the positive direction.
Although the spin inversion process does not change the instantaneous group velocity, due to the spin-dependent force, the group velocity of the spin-down component gradually becomes different from the one of the spin-up component.
Naturally, at the same time, the spin-up wave-packet and the spin-down one gradually separate in the spatial space.
For the spin-up wave-packet, the direction of its group velocity changes when it hits the boundary of first Brillouin zone, and then it moves back to the anti-crossing region.
Thus, partial spin inversion and beam splitting occurs again.

\begin{figure}[!htp]
\begin{center}
\includegraphics[width=1.0\columnwidth]{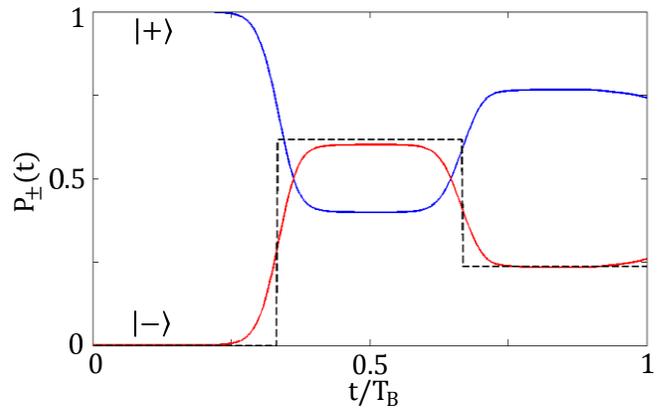}
\end{center}
\caption{(Color online) Non-adiabatic dynamics of dressed-state populations under non-strong inter-spin coupling. The solid lines are populations obtained via numerical calculation. The black dashed line is the population given by the analytical formula~\eqref{LZ+-} and ~\eqref{LZ2}. The parameters are the same as ones used in Fig.~\ref{non-adiabatic-up-down}.}
\label{non_adiabatic_dressed_state}
\end{figure}

One can also understand the non-adiabatic dynamics in the dressed-state picture.
In Fig.~\ref{non_adiabatic_dressed_state}, we show the time evolution of the dressed-state populations.
In the right region far from the anti-crossing region, there is almost no population transfer two dressed states.
Actually, due to large gap in this region, the time-evolution follows the local dressed-`+' states.
In the anti-crossing region, because the upper $V$-shape and the lower $\Lambda$-shape energy ladders are so close, non-adiabatic LZ tunneling between the two energy ladders appears.
Therefore, the population transfer from the dressed-`+' states to the dressed-`-' states takes place.
According to the analytical formula~\eqref{LZ+-}, as the parameters $\hbar\Omega=0.04~E_r$, $F_{m_f}d = 0.01~{E_r}$ and $\hbar v_1=-0.0026~d E_r$, the transition probability for the first inter-spin LZ transition is approximately equal to $P_{-}'=0.617$, which is well consistent with the numerical result $P_{-}=0.602$.
The small relative difference, $\epsilon= \left|P_{-}-P_{-}' \right| / P_{-}=2.5\%$, may be caused by the truncation of Taylor expansion in the analytical derivation.

After the first LZ transition, the wave-packet moves in the left region ($-D_{pi}<jd<0$) and the population transfer between the two dressed states are then suppressed by the large gap.
Therefore the dressed-state populations remain unchanged until the second LZ transition happens and the wave-packet in dressed-`-' states can be approximately regarded as a wave-packet in spin-up state.
In this case, the second LZ transition, which is caused by the return of the spin-up component (dressed-`-' state component) to the anti-crossing region, occurs at time $t_2=T_B-t_1$.
Similar to the analytical formula~\eqref{LZ+-}, one can obtain the transition probability from dressed-`-' states to dressed-`+' states $P_{2,-\rightarrow+}$ in the second LZ transition.
Combining the first and second LZ transitions, the probability of the particle in dressed-`-' states right after the second LZ transition is given by the sequential LZ formula
\begin{equation}
  P_{-}^{'}=P_{1,+\rightarrow-}\left(1-P_{2,-\rightarrow+}\right)=0.236,\label{LZ2}
\end{equation}
which is very close to the numerical result $P_{-}=0.234$.
The relative difference between the analytical and numerical results is $\epsilon=0.9\%$.
The deduction of the relative difference may be caused by the counteraction of truncation errors for the two LZ transitions.

\section{Conclusion and discussion}\label{sec5}

In summary, we have explored the dynamics of single-particle WZ system with inter-spin coupling.
Different from the inter-band LZ tunneling in WS systems, in our WZ system, we explore the novel inter-spin LZ transitions driven by intrinsic intra-band Bloch oscillations, which is called the BLZ dynamics.
If the inter-spin coupling is absent, the wave-packet recovers the conventional Bloch oscillations in the WS systems.
Under strong inter-spin couplings, the spin dynamics adiabatically follows the local dressed states and the wave-packet undergoes joint Bloch-like oscillations of the two spin components.
Under non-strong inter-spin coupling, the inter-spin coupling plays the role of a beam splitter in the spatial anti-crossing region and the two spin components undergo separate Bloch oscillations in the regions away from the spatial anti-crossing.

\begin{figure}[!htp]
\begin{center}
\includegraphics[width=1.0\columnwidth]{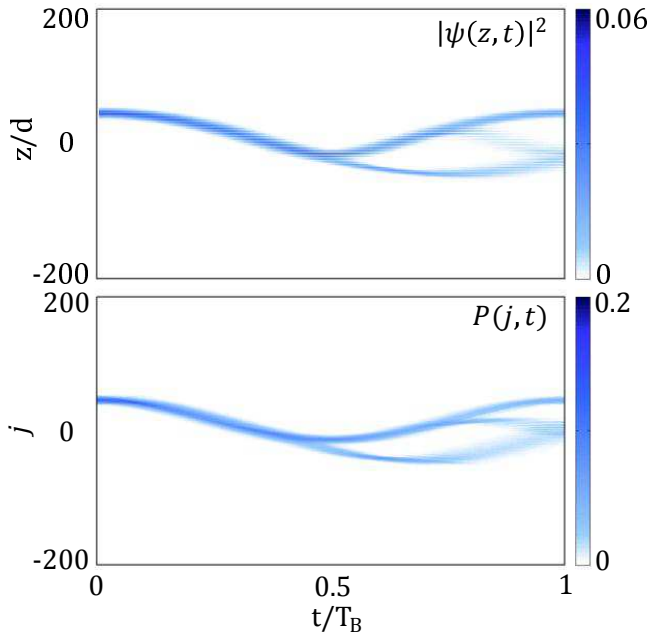}
\end{center}
\caption{(Color online) Validity of the single-band tight-binding approximation.
Top: the dynamics in the original Hamiltonian \eqref{equ:origin} with $V_0 =1.8 E_r$, $\hbar \Omega = 0.04 E_r$, $F_{m_f} d = 0.01E_r$.
Bottom: the dynamics in the corresponding single-band tight-binding Hamiltonian \eqref{equ:single-tight}.}
\label{compare1}
\end{figure}
\begin{figure}[!htp]
\begin{center}
\includegraphics[width=1.0\columnwidth]{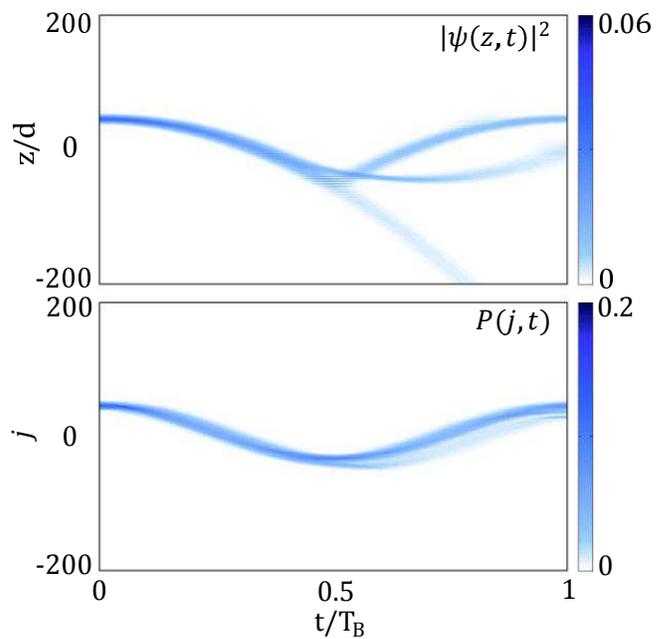}
\end{center}
\caption{(Color online) Breakdown of the single-band tight-binding approximation.
Top: the dynamics of the original Hamiltonian \eqref{equ:origin} with $ V_0 =1.8 E_r $, $ \hbar \Omega = 0.04 E_r $, $ F_{m_f} d = 0.01E_r $.
Bottom: the dynamics of the corresponding single-band tight-binding Hamiltonian~\eqref{equ:single-tight}.}
\label{compare2}
\end{figure}

In this article, we concentrate on the short-time dynamics in single-particle WZ systems with inter-spin coupling.
However, in long-time dynamics, there may appear sequential inter-spin LZ transitions driven by intrinsic intra-band Bloch oscillations.
The sequential inter-spin LZ transitions in long-time dynamics can be regarded as a spatial counterpart for Landau-Zener-St\"{u}ckelberg interferences in two-level system under periodic driving fields~\cite{Ashhab2007,Shevchenko2010}, in which the intrinsic intra-band Bloch oscillations take the role of the periodic driving fields.
By introducing inter-spin coupling into the optical lattices with gradient magnetic field, our WZ system can be realized by current technology~\cite{Kennedy2013,Miyake2013,Aidelsburger2013}.
As the spin-dependent tunneling can be induced by the gradient magnetic field, our single-particle results would be helpful for understanding the spin-orbit coupling effects in one-dimensional quantum lattice systems~\cite{Kennedy2013}.

In addition, the spin-dependent force and the inter-spin coupling in our WZ system are time-independent.
More recently, it has been proposed that artificial spin-orbit coupling can be emulated by applying a periodic gradient magnetic field to a WZ system~\cite{Struck2014}.
Further more, by applying periodic gradient magnetic field and time-dependent inter-spin coupling, it is possible to simulate quantum walks and Dirac dynamics~\cite{PhysRevA.82.033602}.
The time-modulation of the WZ system is an interesting topic deserved further study.

In the end, we would like to discuss briefly the validity of the single-band tight-binding Hamiltonian.
To show the validity of the single-band tight-binding Hamiltonian, we simulate the time evolution of the original Hamiltonian \eqref{equ:origin} (view ~\cite{Feit1982} for the spectral methods) and the corresponding single-band tight-binding Hamiltonian~\eqref{equ:single-tight} under same conditions.
In our above calculations, as the optical lattice potential is sufficiently deep, the dynamics of the original Hamiltonian \eqref{equ:origin} and the corresponding single-band tight-binding Hamiltonian~\eqref{equ:single-tight} almost have no difference, see Fig.~\ref{compare1}.
This means that inter-band tunneling can be ignored and the single-band tight-binding approximation works very well.
If the lattice depth is not large enough, the difference between dynamics of the two Hamiltonians is very significant, see Fig.~\ref{compare2}.
In the original Hamiltonian \eqref{equ:origin}, there appears significant tunneling from the first band to the second band, see the top panel of Fig.~\ref{compare2}.
Therefore, the inter-band LZ transitions and the inter-spin LZ transitions coexist and the single-band tight-binding approximation becomes invlaid.

\acknowledgments
This work is supported by the National Basic Research Program of China (NBRPC) under Grant No. 2012CB821305, the National Natural Science Foundation of China (NNSFC) under Grants No. 11374375, and the PhD Programs Foundation of Ministry of Education of China under Grant No. 20120171110022.

\bibliography{BZL}

\begin{thebibliography}{43}
\expandafter\ifx\csname natexlab\endcsname\relax\def\natexlab#1{#1}\fi
\expandafter\ifx\csname bibnamefont\endcsname\relax
  \def\bibnamefont#1{#1}\fi
\expandafter\ifx\csname bibfnamefont\endcsname\relax
  \def\bibfnamefont#1{#1}\fi
\expandafter\ifx\csname citenamefont\endcsname\relax
  \def\citenamefont#1{#1}\fi
\expandafter\ifx\csname url\endcsname\relax
  \def\url#1{\texttt{#1}}\fi
\expandafter\ifx\csname urlprefix\endcsname\relax\def\urlprefix{URL }\fi
\providecommand{\bibinfo}[2]{#2}
\providecommand{\eprint}[2][]{\url{#2}}

\bibitem[{\citenamefont{Nenciu}(1991)}]{Nenciu1991}
\bibinfo{author}{\bibfnamefont{G.}~\bibnamefont{Nenciu}},
  \bibinfo{journal}{Rev. Mod. Phys.} \textbf{\bibinfo{volume}{63}},
  \bibinfo{pages}{91} (\bibinfo{year}{1991}).

\bibitem[{\citenamefont{Raizen et~al.}(1997)\citenamefont{Raizen, Salomon, and
  Qian}}]{Raizen1997}
\bibinfo{author}{\bibfnamefont{M.}~\bibnamefont{Raizen}},
  \bibinfo{author}{\bibfnamefont{C.}~\bibnamefont{Salomon}}, \bibnamefont{and}
  \bibinfo{author}{\bibfnamefont{N.}~\bibnamefont{Qian}},
  \bibinfo{journal}{Physics Today} \textbf{\bibinfo{volume}{50}},
  \bibinfo{pages}{30} (\bibinfo{year}{1997}).

\bibitem[{\citenamefont{Gl\"{u}ck et~al.}(2002)\citenamefont{Gl\"{u}ck,
  Kolovsky, and Korsch}}]{Korsch2002}
\bibinfo{author}{\bibfnamefont{M.}~\bibnamefont{Gl\"{u}ck}},
  \bibinfo{author}{\bibfnamefont{A.~R.} \bibnamefont{Kolovsky}},
  \bibnamefont{and} \bibinfo{author}{\bibfnamefont{H.~J.}
  \bibnamefont{Korsch}}, \bibinfo{journal}{Physics Reports}
  \textbf{\bibinfo{volume}{366}}, \bibinfo{pages}{103 } (\bibinfo{year}{2002}).

\bibitem[{\citenamefont{Xiao et~al.}(2010)\citenamefont{Xiao, Chang, and
  Niu}}]{Xiao2010}
\bibinfo{author}{\bibfnamefont{D.}~\bibnamefont{Xiao}},
  \bibinfo{author}{\bibfnamefont{M.-C.} \bibnamefont{Chang}}, \bibnamefont{and}
  \bibinfo{author}{\bibfnamefont{Q.}~\bibnamefont{Niu}}, \bibinfo{journal}{Rev.
  Mod. Phys.} \textbf{\bibinfo{volume}{82}}, \bibinfo{pages}{1959}
  (\bibinfo{year}{2010}).

\bibitem[{\citenamefont{Bloch}(1929)}]{Bloch1929}
\bibinfo{author}{\bibfnamefont{F.}~\bibnamefont{Bloch}},
  \bibinfo{journal}{Zeitschrift f¨¹r Physik} \textbf{\bibinfo{volume}{52}},
  \bibinfo{pages}{555} (\bibinfo{year}{1929}).

\bibitem[{\citenamefont{Zener}(1934)}]{Zener1934}
\bibinfo{author}{\bibfnamefont{C.}~\bibnamefont{Zener}},
  \bibinfo{journal}{Proceedings of the Royal Society of London. Series A}
  \textbf{\bibinfo{volume}{145}}, \bibinfo{pages}{523} (\bibinfo{year}{1934}).

\bibitem[{\citenamefont{Wannier}(1960)}]{Wannier1960}
\bibinfo{author}{\bibfnamefont{G.~H.} \bibnamefont{Wannier}},
  \bibinfo{journal}{Phys. Rev.} \textbf{\bibinfo{volume}{117}},
  \bibinfo{pages}{432} (\bibinfo{year}{1960}).

\bibitem[{\citenamefont{Mendez et~al.}(1988)\citenamefont{Mendez,
  Agull\'o-Rueda, and Hong}}]{Mendez1988}
\bibinfo{author}{\bibfnamefont{E.~E.} \bibnamefont{Mendez}},
  \bibinfo{author}{\bibfnamefont{F.}~\bibnamefont{Agull\'o-Rueda}},
  \bibnamefont{and} \bibinfo{author}{\bibfnamefont{J.~M.} \bibnamefont{Hong}},
  \bibinfo{journal}{Phys. Rev. Lett.} \textbf{\bibinfo{volume}{60}},
  \bibinfo{pages}{2426} (\bibinfo{year}{1988}).

\bibitem[{\citenamefont{Waschke et~al.}(1993)\citenamefont{Waschke, Roskos,
  Schwedler, Leo, Kurz, and K\"ohler}}]{Waschke1993}
\bibinfo{author}{\bibfnamefont{C.}~\bibnamefont{Waschke}},
  \bibinfo{author}{\bibfnamefont{H.~G.} \bibnamefont{Roskos}},
  \bibinfo{author}{\bibfnamefont{R.}~\bibnamefont{Schwedler}},
  \bibinfo{author}{\bibfnamefont{K.}~\bibnamefont{Leo}},
  \bibinfo{author}{\bibfnamefont{H.}~\bibnamefont{Kurz}}, \bibnamefont{and}
  \bibinfo{author}{\bibfnamefont{K.}~\bibnamefont{K\"ohler}},
  \bibinfo{journal}{Phys. Rev. Lett.} \textbf{\bibinfo{volume}{70}},
  \bibinfo{pages}{3319} (\bibinfo{year}{1993}).

\bibitem[{\citenamefont{Dekorsy et~al.}(1994)\citenamefont{Dekorsy, Leisching,
  K\"ohler, and Kurz}}]{Dekorsy1994}
\bibinfo{author}{\bibfnamefont{T.}~\bibnamefont{Dekorsy}},
  \bibinfo{author}{\bibfnamefont{P.}~\bibnamefont{Leisching}},
  \bibinfo{author}{\bibfnamefont{K.}~\bibnamefont{K\"ohler}}, \bibnamefont{and}
  \bibinfo{author}{\bibfnamefont{H.}~\bibnamefont{Kurz}},
  \bibinfo{journal}{Phys. Rev. B} \textbf{\bibinfo{volume}{50}},
  \bibinfo{pages}{8106} (\bibinfo{year}{1994}).

\bibitem[{\citenamefont{Lyssenko et~al.}(1997)\citenamefont{Lyssenko,
  Valu\ifmmode~\check{s}\else \v{s}\fi{}is, L\"oser, Hasche, Leo, Dignam, and
  K\"ohler}}]{Lyssenko1997}
\bibinfo{author}{\bibfnamefont{V.~G.} \bibnamefont{Lyssenko}},
  \bibinfo{author}{\bibfnamefont{G.}~\bibnamefont{Valu\ifmmode~\check{s}\else
  \v{s}\fi{}is}}, \bibinfo{author}{\bibfnamefont{F.}~\bibnamefont{L\"oser}},
  \bibinfo{author}{\bibfnamefont{T.}~\bibnamefont{Hasche}},
  \bibinfo{author}{\bibfnamefont{K.}~\bibnamefont{Leo}},
  \bibinfo{author}{\bibfnamefont{M.~M.} \bibnamefont{Dignam}},
  \bibnamefont{and} \bibinfo{author}{\bibfnamefont{K.}~\bibnamefont{K\"ohler}},
  \bibinfo{journal}{Phys. Rev. Lett.} \textbf{\bibinfo{volume}{79}},
  \bibinfo{pages}{301} (\bibinfo{year}{1997}).

\bibitem[{\citenamefont{Peschel et~al.}(1998)\citenamefont{Peschel, Pertsch,
  and Lederer}}]{Peschel1998}
\bibinfo{author}{\bibfnamefont{U.}~\bibnamefont{Peschel}},
  \bibinfo{author}{\bibfnamefont{T.}~\bibnamefont{Pertsch}}, \bibnamefont{and}
  \bibinfo{author}{\bibfnamefont{F.}~\bibnamefont{Lederer}},
  \bibinfo{journal}{Opt. Lett.} \textbf{\bibinfo{volume}{23}},
  \bibinfo{pages}{1701} (\bibinfo{year}{1998}).

\bibitem[{\citenamefont{Pertsch et~al.}(1999)\citenamefont{Pertsch, Dannberg,
  Elflein, Br\"auer, and Lederer}}]{Pertsch1999}
\bibinfo{author}{\bibfnamefont{T.}~\bibnamefont{Pertsch}},
  \bibinfo{author}{\bibfnamefont{P.}~\bibnamefont{Dannberg}},
  \bibinfo{author}{\bibfnamefont{W.}~\bibnamefont{Elflein}},
  \bibinfo{author}{\bibfnamefont{A.}~\bibnamefont{Br\"auer}}, \bibnamefont{and}
  \bibinfo{author}{\bibfnamefont{F.}~\bibnamefont{Lederer}},
  \bibinfo{journal}{Phys. Rev. Lett.} \textbf{\bibinfo{volume}{83}},
  \bibinfo{pages}{4752} (\bibinfo{year}{1999}).

\bibitem[{\citenamefont{Morandotti et~al.}(1999)\citenamefont{Morandotti,
  Peschel, Aitchison, Eisenberg, and Silberberg}}]{Morandotti1999}
\bibinfo{author}{\bibfnamefont{R.}~\bibnamefont{Morandotti}},
  \bibinfo{author}{\bibfnamefont{U.}~\bibnamefont{Peschel}},
  \bibinfo{author}{\bibfnamefont{J.~S.} \bibnamefont{Aitchison}},
  \bibinfo{author}{\bibfnamefont{H.~S.} \bibnamefont{Eisenberg}},
  \bibnamefont{and}
  \bibinfo{author}{\bibfnamefont{Y.}~\bibnamefont{Silberberg}},
  \bibinfo{journal}{Phys. Rev. Lett.} \textbf{\bibinfo{volume}{83}},
  \bibinfo{pages}{4756} (\bibinfo{year}{1999}).

\bibitem[{\citenamefont{Pertsch et~al.}(2002)\citenamefont{Pertsch, Zentgraf,
  Peschel, Br\"auer, and Lederer}}]{Pertsch2002}
\bibinfo{author}{\bibfnamefont{T.}~\bibnamefont{Pertsch}},
  \bibinfo{author}{\bibfnamefont{T.}~\bibnamefont{Zentgraf}},
  \bibinfo{author}{\bibfnamefont{U.}~\bibnamefont{Peschel}},
  \bibinfo{author}{\bibfnamefont{A.}~\bibnamefont{Br\"auer}}, \bibnamefont{and}
  \bibinfo{author}{\bibfnamefont{F.}~\bibnamefont{Lederer}},
  \bibinfo{journal}{Phys. Rev. Lett.} \textbf{\bibinfo{volume}{88}},
  \bibinfo{pages}{093901} (\bibinfo{year}{2002}).

\bibitem[{\citenamefont{Ben~Dahan et~al.}(1996)\citenamefont{Ben~Dahan, Peik,
  Reichel, Castin, and Salomon}}]{Dahan1996}
\bibinfo{author}{\bibfnamefont{M.}~\bibnamefont{Ben~Dahan}},
  \bibinfo{author}{\bibfnamefont{E.}~\bibnamefont{Peik}},
  \bibinfo{author}{\bibfnamefont{J.}~\bibnamefont{Reichel}},
  \bibinfo{author}{\bibfnamefont{Y.}~\bibnamefont{Castin}}, \bibnamefont{and}
  \bibinfo{author}{\bibfnamefont{C.}~\bibnamefont{Salomon}},
  \bibinfo{journal}{Phys. Rev. Lett.} \textbf{\bibinfo{volume}{76}},
  \bibinfo{pages}{4508} (\bibinfo{year}{1996}).

\bibitem[{\citenamefont{Niu et~al.}(1996)\citenamefont{Niu, Zhao, Georgakis,
  and Raizen}}]{Niu1996}
\bibinfo{author}{\bibfnamefont{Q.}~\bibnamefont{Niu}},
  \bibinfo{author}{\bibfnamefont{X.-G.} \bibnamefont{Zhao}},
  \bibinfo{author}{\bibfnamefont{G.~A.} \bibnamefont{Georgakis}},
  \bibnamefont{and} \bibinfo{author}{\bibfnamefont{M.~G.}
  \bibnamefont{Raizen}}, \bibinfo{journal}{Phys. Rev. Lett.}
  \textbf{\bibinfo{volume}{76}}, \bibinfo{pages}{4504} (\bibinfo{year}{1996}).

\bibitem[{\citenamefont{Wilkinson et~al.}(1996)\citenamefont{Wilkinson,
  Bharucha, Madison, Niu, and Raizen}}]{Wilkinson1996}
\bibinfo{author}{\bibfnamefont{S.~R.} \bibnamefont{Wilkinson}},
  \bibinfo{author}{\bibfnamefont{C.~F.} \bibnamefont{Bharucha}},
  \bibinfo{author}{\bibfnamefont{K.~W.} \bibnamefont{Madison}},
  \bibinfo{author}{\bibfnamefont{Q.}~\bibnamefont{Niu}}, \bibnamefont{and}
  \bibinfo{author}{\bibfnamefont{M.~G.} \bibnamefont{Raizen}},
  \bibinfo{journal}{Phys. Rev. Lett.} \textbf{\bibinfo{volume}{76}},
  \bibinfo{pages}{4512} (\bibinfo{year}{1996}).

\bibitem[{\citenamefont{Peik et~al.}(1997)\citenamefont{Peik, Ben~Dahan,
  Bouchoule, Castin, and Salomon}}]{Peik1997}
\bibinfo{author}{\bibfnamefont{E.}~\bibnamefont{Peik}},
  \bibinfo{author}{\bibfnamefont{M.}~\bibnamefont{Ben~Dahan}},
  \bibinfo{author}{\bibfnamefont{I.}~\bibnamefont{Bouchoule}},
  \bibinfo{author}{\bibfnamefont{Y.}~\bibnamefont{Castin}}, \bibnamefont{and}
  \bibinfo{author}{\bibfnamefont{C.}~\bibnamefont{Salomon}},
  \bibinfo{journal}{Phys. Rev. A} \textbf{\bibinfo{volume}{55}},
  \bibinfo{pages}{2989} (\bibinfo{year}{1997}).

\bibitem[{\citenamefont{Bharucha et~al.}(1997)\citenamefont{Bharucha, Madison,
  Morrow, Wilkinson, Sundaram, and Raizen}}]{Bharucha1997}
\bibinfo{author}{\bibfnamefont{C.~F.} \bibnamefont{Bharucha}},
  \bibinfo{author}{\bibfnamefont{K.~W.} \bibnamefont{Madison}},
  \bibinfo{author}{\bibfnamefont{P.~R.} \bibnamefont{Morrow}},
  \bibinfo{author}{\bibfnamefont{S.~R.} \bibnamefont{Wilkinson}},
  \bibinfo{author}{\bibfnamefont{B.}~\bibnamefont{Sundaram}}, \bibnamefont{and}
  \bibinfo{author}{\bibfnamefont{M.~G.} \bibnamefont{Raizen}},
  \bibinfo{journal}{Phys. Rev. A} \textbf{\bibinfo{volume}{55}},
  \bibinfo{pages}{R857} (\bibinfo{year}{1997}).

\bibitem[{\citenamefont{Morsch et~al.}(2001)\citenamefont{Morsch, M\"uller,
  Cristiani, Ciampini, and Arimondo}}]{Morsch2001}
\bibinfo{author}{\bibfnamefont{O.}~\bibnamefont{Morsch}},
  \bibinfo{author}{\bibfnamefont{J.~H.} \bibnamefont{M\"uller}},
  \bibinfo{author}{\bibfnamefont{M.}~\bibnamefont{Cristiani}},
  \bibinfo{author}{\bibfnamefont{D.}~\bibnamefont{Ciampini}}, \bibnamefont{and}
  \bibinfo{author}{\bibfnamefont{E.}~\bibnamefont{Arimondo}},
  \bibinfo{journal}{Phys. Rev. Lett.} \textbf{\bibinfo{volume}{87}},
  \bibinfo{pages}{140402} (\bibinfo{year}{2001}).

\bibitem[{\citenamefont{Atala et~al.}(2013)\citenamefont{Atala, Aidelsburger,
  Barreiro, Abanin, Kitagawa, Demler, and Bloch}}]{Atala2013}
\bibinfo{author}{\bibfnamefont{M.}~\bibnamefont{Atala}},
  \bibinfo{author}{\bibfnamefont{M.}~\bibnamefont{Aidelsburger}},
  \bibinfo{author}{\bibfnamefont{J.~T.} \bibnamefont{Barreiro}},
  \bibinfo{author}{\bibfnamefont{D.}~\bibnamefont{Abanin}},
  \bibinfo{author}{\bibfnamefont{T.}~\bibnamefont{Kitagawa}},
  \bibinfo{author}{\bibfnamefont{E.}~\bibnamefont{Demler}}, \bibnamefont{and}
  \bibinfo{author}{\bibfnamefont{I.}~\bibnamefont{Bloch}},
  \bibinfo{journal}{Nature Physics} \textbf{\bibinfo{volume}{9}},
  \bibinfo{pages}{795} (\bibinfo{year}{2013}).

\bibitem[{\citenamefont{{Kolovsky} and {Mantica}}(2014)}]{Kolovsky2014}
\bibinfo{author}{\bibfnamefont{A.~R.} \bibnamefont{{Kolovsky}}}
  \bibnamefont{and}
  \bibinfo{author}{\bibfnamefont{G.}~\bibnamefont{{Mantica}}},
  \bibinfo{journal}{ArXiv e-prints}  (\bibinfo{year}{2014}),
  \eprint{1406.0276}.

\bibitem[{\citenamefont{{Clad{\'e}}}(2014)}]{Clade2014}
\bibinfo{author}{\bibfnamefont{P.}~\bibnamefont{{Clad{\'e}}}},
  \bibinfo{journal}{ArXiv e-prints}  (\bibinfo{year}{2014}),
  \eprint{1405.2770}.

\bibitem[{\citenamefont{Hartmann et~al.}(2004)\citenamefont{Hartmann, Keck,
  Korsch, and Mossmann}}]{Hartmann2004}
\bibinfo{author}{\bibfnamefont{T.}~\bibnamefont{Hartmann}},
  \bibinfo{author}{\bibfnamefont{F.}~\bibnamefont{Keck}},
  \bibinfo{author}{\bibfnamefont{H.~J.} \bibnamefont{Korsch}},
  \bibnamefont{and} \bibinfo{author}{\bibfnamefont{S.}~\bibnamefont{Mossmann}},
  \bibinfo{journal}{New J. Phys.} \textbf{\bibinfo{volume}{6}},
  \bibinfo{pages}{2} (\bibinfo{year}{2004}).

\bibitem[{\citenamefont{Holthaus and Hone}(1996)}]{Holthaus1996}
\bibinfo{author}{\bibfnamefont{M.}~\bibnamefont{Holthaus}} \bibnamefont{and}
  \bibinfo{author}{\bibfnamefont{D.~W.} \bibnamefont{Hone}},
  \bibinfo{journal}{Philosophical Magazine Part B}
  \textbf{\bibinfo{volume}{74}}, \bibinfo{pages}{105} (\bibinfo{year}{1996}).

\bibitem[{\citenamefont{Breid et~al.}(2006)\citenamefont{Breid, Witthaut, and
  Korsch}}]{Breid2006}
\bibinfo{author}{\bibfnamefont{B.~M.} \bibnamefont{Breid}},
  \bibinfo{author}{\bibfnamefont{D.}~\bibnamefont{Witthaut}}, \bibnamefont{and}
  \bibinfo{author}{\bibfnamefont{H.~J.} \bibnamefont{Korsch}},
  \bibinfo{journal}{New J. Phys.} \textbf{\bibinfo{volume}{8}},
  \bibinfo{pages}{110} (\bibinfo{year}{2006}).

\bibitem[{\citenamefont{Dreisow et~al.}(2009)\citenamefont{Dreisow, Szameit,
  Heinrich, Pertsch, Nolte, T\"unnermann, and Longhi}}]{Dreisow2009}
\bibinfo{author}{\bibfnamefont{F.}~\bibnamefont{Dreisow}},
  \bibinfo{author}{\bibfnamefont{A.}~\bibnamefont{Szameit}},
  \bibinfo{author}{\bibfnamefont{M.}~\bibnamefont{Heinrich}},
  \bibinfo{author}{\bibfnamefont{T.}~\bibnamefont{Pertsch}},
  \bibinfo{author}{\bibfnamefont{S.}~\bibnamefont{Nolte}},
  \bibinfo{author}{\bibfnamefont{A.}~\bibnamefont{T\"unnermann}},
  \bibnamefont{and} \bibinfo{author}{\bibfnamefont{S.}~\bibnamefont{Longhi}},
  \bibinfo{journal}{Phys. Rev. Lett.} \textbf{\bibinfo{volume}{102}},
  \bibinfo{pages}{076802} (\bibinfo{year}{2009}).

\bibitem[{\citenamefont{Witthaut}(2010)}]{PhysRevA.82.033602}
\bibinfo{author}{\bibfnamefont{D.}~\bibnamefont{Witthaut}},
  \bibinfo{journal}{Phys. Rev. A} \textbf{\bibinfo{volume}{82}},
  \bibinfo{pages}{033602} (\bibinfo{year}{2010}).

\bibitem[{\citenamefont{Kennedy et~al.}(2013)\citenamefont{Kennedy, Siviloglou,
  Miyake, Burton, and Ketterle}}]{Kennedy2013}
\bibinfo{author}{\bibfnamefont{C.~J.} \bibnamefont{Kennedy}},
  \bibinfo{author}{\bibfnamefont{G.~A.} \bibnamefont{Siviloglou}},
  \bibinfo{author}{\bibfnamefont{H.}~\bibnamefont{Miyake}},
  \bibinfo{author}{\bibfnamefont{W.~C.} \bibnamefont{Burton}},
  \bibnamefont{and} \bibinfo{author}{\bibfnamefont{W.}~\bibnamefont{Ketterle}},
  \bibinfo{journal}{Phys. Rev. Lett.} \textbf{\bibinfo{volume}{111}},
  \bibinfo{pages}{225301} (\bibinfo{year}{2013}).

\bibitem[{\citenamefont{Miyake et~al.}(2013)\citenamefont{Miyake, Siviloglou,
  Kennedy, Burton, and Ketterle}}]{Miyake2013}
\bibinfo{author}{\bibfnamefont{H.}~\bibnamefont{Miyake}},
  \bibinfo{author}{\bibfnamefont{G.~A.} \bibnamefont{Siviloglou}},
  \bibinfo{author}{\bibfnamefont{C.~J.} \bibnamefont{Kennedy}},
  \bibinfo{author}{\bibfnamefont{W.~C.} \bibnamefont{Burton}},
  \bibnamefont{and} \bibinfo{author}{\bibfnamefont{W.}~\bibnamefont{Ketterle}},
  \bibinfo{journal}{Phys. Rev. Lett.} \textbf{\bibinfo{volume}{111}},
  \bibinfo{pages}{185302} (\bibinfo{year}{2013}).

\bibitem[{\citenamefont{Aidelsburger et~al.}(2013)\citenamefont{Aidelsburger,
  Atala, Lohse, Barreiro, Paredes, and Bloch}}]{Aidelsburger2013}
\bibinfo{author}{\bibfnamefont{M.}~\bibnamefont{Aidelsburger}},
  \bibinfo{author}{\bibfnamefont{M.}~\bibnamefont{Atala}},
  \bibinfo{author}{\bibfnamefont{M.}~\bibnamefont{Lohse}},
  \bibinfo{author}{\bibfnamefont{J.~T.} \bibnamefont{Barreiro}},
  \bibinfo{author}{\bibfnamefont{B.}~\bibnamefont{Paredes}}, \bibnamefont{and}
  \bibinfo{author}{\bibfnamefont{I.}~\bibnamefont{Bloch}},
  \bibinfo{journal}{Phys. Rev. Lett.} \textbf{\bibinfo{volume}{111}},
  \bibinfo{pages}{185301} (\bibinfo{year}{2013}).

\bibitem[{\citenamefont{Bloch et~al.}(2008)\citenamefont{Bloch, Dalibard, and
  Zwerger}}]{Bloch2008}
\bibinfo{author}{\bibfnamefont{I.}~\bibnamefont{Bloch}},
  \bibinfo{author}{\bibfnamefont{J.}~\bibnamefont{Dalibard}}, \bibnamefont{and}
  \bibinfo{author}{\bibfnamefont{W.}~\bibnamefont{Zwerger}},
  \bibinfo{journal}{Rev. Mod. Phys.} \textbf{\bibinfo{volume}{80}},
  \bibinfo{pages}{885} (\bibinfo{year}{2008}).

\bibitem[{\citenamefont{Morsch and Oberthaler}(2006)}]{Morsch2006}
\bibinfo{author}{\bibfnamefont{O.}~\bibnamefont{Morsch}} \bibnamefont{and}
  \bibinfo{author}{\bibfnamefont{M.}~\bibnamefont{Oberthaler}},
  \bibinfo{journal}{Rev. Mod. Phys.} \textbf{\bibinfo{volume}{78}},
  \bibinfo{pages}{179} (\bibinfo{year}{2006}).

\bibitem[{\citenamefont{Pethick and Smith}(2008)}]{Pethick2008}
\bibinfo{author}{\bibfnamefont{C.}~\bibnamefont{Pethick}} \bibnamefont{and}
  \bibinfo{author}{\bibfnamefont{H.}~\bibnamefont{Smith}},
  \emph{\bibinfo{title}{Bose-Einstein condensation in dilute gases}}
  (\bibinfo{publisher}{Cambridge University Press}, \bibinfo{year}{2008}),
  \bibinfo{edition}{2nd} ed.

\bibitem[{\citenamefont{Walters et~al.}(2013)\citenamefont{Walters, Cotugno,
  Johnson, Clark, and Jaksch}}]{Walters2013}
\bibinfo{author}{\bibfnamefont{R.}~\bibnamefont{Walters}},
  \bibinfo{author}{\bibfnamefont{G.}~\bibnamefont{Cotugno}},
  \bibinfo{author}{\bibfnamefont{T.~H.} \bibnamefont{Johnson}},
  \bibinfo{author}{\bibfnamefont{S.~R.} \bibnamefont{Clark}}, \bibnamefont{and}
  \bibinfo{author}{\bibfnamefont{D.}~\bibnamefont{Jaksch}},
  \bibinfo{journal}{Phys. Rev. A} \textbf{\bibinfo{volume}{87}},
  \bibinfo{pages}{043613} (\bibinfo{year}{2013}).

\bibitem[{\citenamefont{Mostofi et~al.}(2008)\citenamefont{Mostofi, Yates, Lee,
  Souza, Vanderbilt, and Marzari}}]{Arash2008}
\bibinfo{author}{\bibfnamefont{A.~A.} \bibnamefont{Mostofi}},
  \bibinfo{author}{\bibfnamefont{J.~R.} \bibnamefont{Yates}},
  \bibinfo{author}{\bibfnamefont{Y.-S.} \bibnamefont{Lee}},
  \bibinfo{author}{\bibfnamefont{I.}~\bibnamefont{Souza}},
  \bibinfo{author}{\bibfnamefont{D.}~\bibnamefont{Vanderbilt}},
  \bibnamefont{and} \bibinfo{author}{\bibfnamefont{N.}~\bibnamefont{Marzari}},
  \bibinfo{journal}{Computer Physics Communications}
  \textbf{\bibinfo{volume}{178}}, \bibinfo{pages}{685 } (\bibinfo{year}{2008}).

\bibitem[{\citenamefont{Landau and Lifshitz}(1981)}]{Landau1981Quantum}
\bibinfo{author}{\bibfnamefont{L.~D.} \bibnamefont{Landau}} \bibnamefont{and}
  \bibinfo{author}{\bibfnamefont{L.~M.} \bibnamefont{Lifshitz}},
  \emph{\bibinfo{title}{Quantum Mechanics Non-Relativistic Theory, Third
  Edition: Volume 3}} (\bibinfo{publisher}{Butterworth-Heinemann},
  \bibinfo{year}{1981}), \bibinfo{edition}{3rd} ed., ISBN
  \bibinfo{isbn}{0750635398}.

\bibitem[{\citenamefont{Ashhab et~al.}(2007)\citenamefont{Ashhab, Johansson,
  Zagoskin, and Nori}}]{Ashhab2007}
\bibinfo{author}{\bibfnamefont{S.}~\bibnamefont{Ashhab}},
  \bibinfo{author}{\bibfnamefont{J.~R.} \bibnamefont{Johansson}},
  \bibinfo{author}{\bibfnamefont{A.~M.} \bibnamefont{Zagoskin}},
  \bibnamefont{and} \bibinfo{author}{\bibfnamefont{F.}~\bibnamefont{Nori}},
  \bibinfo{journal}{Phys. Rev. A} \textbf{\bibinfo{volume}{75}},
  \bibinfo{pages}{063414} (\bibinfo{year}{2007}).

\bibitem[{\citenamefont{Shevchenko et~al.}(2010)\citenamefont{Shevchenko,
  Ashhab, and Nori}}]{Shevchenko2010}
\bibinfo{author}{\bibfnamefont{S.}~\bibnamefont{Shevchenko}},
  \bibinfo{author}{\bibfnamefont{S.}~\bibnamefont{Ashhab}}, \bibnamefont{and}
  \bibinfo{author}{\bibfnamefont{F.}~\bibnamefont{Nori}},
  \bibinfo{journal}{Physics Reports} \textbf{\bibinfo{volume}{492}},
  \bibinfo{pages}{1 } (\bibinfo{year}{2010}).

\bibitem[{\citenamefont{Struck et~al.}(2014)\citenamefont{Struck, Simonet, and
  Sengstock}}]{Struck2014}
\bibinfo{author}{\bibfnamefont{J.}~\bibnamefont{Struck}},
  \bibinfo{author}{\bibfnamefont{J.}~\bibnamefont{Simonet}}, \bibnamefont{and}
  \bibinfo{author}{\bibfnamefont{K.}~\bibnamefont{Sengstock}},
  \bibinfo{journal}{Phys. Rev. A} \textbf{\bibinfo{volume}{90}},
  \bibinfo{pages}{031601} (\bibinfo{year}{2014}).

\bibitem[{\citenamefont{Feit et~al.}(1982)\citenamefont{Feit, Jr., and
  Steiger}}]{Feit1982}
\bibinfo{author}{\bibfnamefont{M.}~\bibnamefont{Feit}},
  \bibinfo{author}{\bibfnamefont{J.~F.} \bibnamefont{Jr.}}, \bibnamefont{and}
  \bibinfo{author}{\bibfnamefont{A.}~\bibnamefont{Steiger}},
  \bibinfo{journal}{Journal of Computational Physics}
  \textbf{\bibinfo{volume}{47}}, \bibinfo{pages}{412 } (\bibinfo{year}{1982}).

\bibitem[{\citenamefont{Essler et~al.}(2005)\citenamefont{Essler, Frahm,
  G{\"o}hmann, Kl{\"u}mper, and Korepin}}]{Fabian2005}
\bibinfo{author}{\bibfnamefont{F.~H.~L.} \bibnamefont{Essler}},
  \bibinfo{author}{\bibfnamefont{H.}~\bibnamefont{Frahm}},
  \bibinfo{author}{\bibfnamefont{F.}~\bibnamefont{G{\"o}hmann}},
  \bibinfo{author}{\bibfnamefont{A.}~\bibnamefont{Kl{\"u}mper}},
  \bibnamefont{and} \bibinfo{author}{\bibfnamefont{V.~E.}
  \bibnamefont{Korepin}}, \emph{\bibinfo{title}{The One-Dimensional {H}ubbard
  Model}} (\bibinfo{publisher}{Cambridge University Press},
  \bibinfo{address}{Cambridge}, \bibinfo{year}{2005}).

\end{thebibliography}

\appendix

\section{Derivation of the single-band tight-binding Hamiltonian}

In this appendix, we show how to derive the single-band tight-binding Hamiltonian \eqref{equ:single-tight}.

The wavefunctions of the system~\eqref{equ:origin} can be written as $\psi= \sum_s {\varphi_s(z)\chi(s)}$, where $\varphi_s(z)$ describe the external spatial states and $\chi(s)$ denotes the internal spin states. In the $\hat\sigma_z$-representation,
\begin{equation}
  \chi(\uparrow)=\left(
  \begin{matrix}
   1 \\
   0
  \end{matrix}
  \right), \;
    \chi(\downarrow)=\left(
  \begin{matrix}
   0 \\
   1
  \end{matrix}
  \right).
\end{equation}
It is obvious that $\chi^{\dagger}(s')\chi(s)=\delta_{s's}$. According to Bloch's theorem~\cite{Fabian2005}, the eigenequation of $\hat H_0$ reads as
\begin{equation}
\hat H_0 \varphi_{n,q,s}(z)\chi(s)
=\varepsilon_{n,q}\varphi_{n,q,s}(z)\chi(s),
\end{equation}
where the eigenfunctions $\varphi_{n,q,s}(z)$, the so-called Bloch functions, are in form of
\begin{equation}
  \varphi_{n,q,s}(z)=e^{iqz}u_{n,q}(z).
\end{equation}
Here, ${u_{n,q}}(z)$ is a periodic function with the period $d$, the band index $n$ and the quasi-momentum $q$, and the first Brillouin zone is given as $|q|\le2\pi/d$.
In addition, the Bloch functions satisfy with the orthogonality conditions,
\begin{equation}
  \int\varphi_{n',q',s}^*(z)\varphi_{n,q,s}(z)\mathrm{d}z=\delta_{n'n}\delta_{q'q}.
\end{equation}

To clearly show the particle hopping between different lattice sites, one may alternatively choose another set of basis called Wannier functions.
The Bloch functions can be expanded in terms of the Wannier functions by the following transformation,
\begin{equation}
{\varphi _{n,q,s}}\left( z \right) = {1 \over {\sqrt L }}\sum\limits_j^{} {{e^{iq{R_j}}}{\phi _{n,s}}\left( {z - {R_j}} \right)},
\end{equation}
where, ${R_j} = jd$, $L$ is the total lattice number, and the Wannier function ${\phi _{n,s}}\left( {z - {R_j}} \right)$ centers in the $j$-th lattice site.
One can also obtain the Wannier functions from the Bloch functions via the inverse transformation,
\begin{equation}
{\phi _{n,s}}\left( {z - {R_j}} \right) = {1 \over {\sqrt L }}\sum\limits_q^{} {{e^{-iq{R_j}}}{\varphi _{n,q,s}}\left( z \right).}
\end{equation}

Denoting the spin-dependent Wannier basis as $\left|{n,j,s}\right\rangle$, we have $\phi_{n,s}(z-R_j)\chi(s)=\left\langle{{z,s}}\mathrel{\left|{\vphantom{{z}{n,j}}}\right.\kern-\nulldelimiterspace}{{n,j,s}}\right\rangle$,
${\hat{\sigma}_z} \left|{n,j,s}\right\rangle = \mathrm{sgn}(s) \left|{n,j,s}\right\rangle$, and ${\hat{\sigma}_x} \left|{n,j,s}\right\rangle = \left|{n,j,-s}\right\rangle$,
in which we define
$$
\mathrm{sgn}(s)=
\begin{cases}
+1  & \textrm{for~~} s=\uparrow, \\
-1  & \textrm{for~~} s=\downarrow,
\end{cases}
\textrm{~~and~~}
 - s  =
\begin{cases}
\uparrow      &  \textrm{for~~} s = \downarrow, \\
\downarrow    &  \textrm{for~~} s = \uparrow.
\end{cases}
$$
Therefore, the matrix elements of the whole Hamiltonian are given as,
\begin{widetext}
\begin{eqnarray}
\left\langle {m,k,s'} \right|\hat H\left| {n,j,s} \right\rangle  = \left\langle {m,k,s'} \right|{\hat H}_0\left| {n,j,s} \right\rangle + \left\langle {m,k,s'} \right| {\hat H _t}\left| {n,j,s} \right\rangle + \left\langle {m,k,s'} \right|{\hat H _c}\left| {n,j,s} \right\rangle,
\end{eqnarray}
with the first term
\begin{equation}
  \begin{split}
  \left\langle{m,k,s'}\right| {\hat H_0} \left|{n,j,s}\right\rangle
  & = \delta_{s',s}\int_{}^{} {\phi _{m,s'}^*\left( {z - {R_k}} \right) h_0 {\phi _{n,s}}\left( {z - {R_j}} \right)\mathrm{d}z}
   = \delta_{s',s}{1 \over L}\sum\limits_{q',q}^{} {{e^{iq'{R_k}}}{e^{ -iq{R_j}}}\int{\varphi_{m,q',s'}^*\left( z \right){ h_0}} {\varphi _{n,q,s}}\left( z \right)\mathrm{d}z}   \\
  & = \delta_{s',s}\delta_{m,n}{1 \over L}\sum\limits_q^{} {{e^{iq\left( {{R_k} - {R_j}} \right)}}{\varepsilon _{n,q}}},  \\
  \end{split}
\end{equation}
the second term
\begin{equation}
  \begin{split}
  \left\langle {m,k,s'} \right| {\hat{H}_t}\left| {n,j,s} \right\rangle
  & = \mathrm{sgn}(s)\delta_{s',s}\int {\phi _{m,s'}^*\left( {z - {R_k}} \right)F_{m_f}z{\phi _{n,s}}\left( {z - {R_j}} \right)\mathrm{d}z}   \\
  & = \mathrm{sgn}(s)\delta_{s',s}{F_{m_f} \over L}\sum\limits_{q',q}^{} {{e^{iq'{R_k}}}{e^{  -iq{R_j}}}\int_{}^{} {\varphi _{m,q',s'}^*\left( z \right)z{\varphi _{n,q,s}}\left( z \right)\mathrm{d}z} }   \\
  & = \mathrm{sgn}(s)\delta_{s',s}{F_{m_f} \over L}\sum\limits_q^{} {{e^{iq\left( {{R_k} - {R_j}} \right)}}\int{u_{m,q}^*\left( z \right)i{\mathrm{d} \over {\mathrm{d}q}}{u_{n,q}}\left( z \right)\mathrm{d}z} }
  + \mathrm{sgn}(s)\delta_{s',s}F_{m_f}R_j\delta _{m,n}\delta_{j,k}, \\
  \end{split}
\end{equation}
and the last term
\begin{equation}
  \begin{split}
  \left\langle {m,k,s'} \right|{\hat{H}_c}\left| {n,j,s} \right\rangle
  & = \delta_{s',-s}\int{\phi _{m,s'}^*\left( {z - {R_k}} \right){{\hbar \Omega } \over 2}{\phi _{n,-s}}\left( {z - {R_j}} \right)\mathrm{d}z}   \cr
  &  = \delta_{s',-s} {{\hbar \Omega } \over 2}{1 \over L}\sum\limits_{q',q}^{} {{e^{iq'{R_k}}}{e^{-iq{R_j}}}\int_{}^{} {\varphi _{m,q',s'}^*\left( z \right){\varphi _{n,q,-s}}\left( z \right)\mathrm{d}z} }  = {{\hbar \Omega } \over 2}
  \delta_{s',-s}{\delta _{m,n}}{\delta _{k,j}}. \cr
  \end{split}
\end{equation}
Thus the Hamiltonian can be expressed as,
\begin{equation}
\begin{split}
    \hat H = & \sum\limits_{n,\left\langle {j,k} \right\rangle ,s }^{} {t_{j,k}^n\left| {n,j,s} \right\rangle \left\langle {n,k,s} \right|} + F_{m_f}d\sum\limits_{n,j,s}^{} {j{\hat{\sigma} _z}\left| {n,j,s} \right\rangle \left\langle {n,j,s} \right|}   + {{\hbar \Omega } \over 2}\sum\limits_{n,j}^{} \left({\left| { n,j,\downarrow } \right\rangle \left\langle {n,j,\uparrow} \right| + \mathrm{h.c.}} \right)   \cr
   & + \sum\limits_{\left\langle {m,n} \right\rangle ,\left\langle {j,k} \right\rangle ,s}^{} {M_{n,j}^{m,k}{\hat{\sigma}}_z \left| {m,k,s} \right\rangle \left\langle {n,j,s} \right|},  \cr
\end{split}
\end{equation}
in which, the on-site energy $t_{j,j}^n$ and the intra-band hopping strengths $t_{j,k}^n$ ($j \ne k$) read as
\begin{equation}
\label{equ:tunneljk}
t_{j,k}^n = {1 \over L}\sum\limits_q^{} {{e^{iq\left( {{R_k} - {R_j}} \right)}}{\varepsilon _{n,q}},}
\end{equation}
and the inter-band tunneling strengths ${M_{n,j}^{m,k}}$ are given by
\begin{equation}
\begin{split}
 M_{n,j}^{m,k} = {F_{m_f} \over L} \sum\limits_q^{} {{e^{iq\left( {{R_k} - {R_j}} \right)}}\int_{}^{} {u_{m,q}^*\left(z \right)i{\mathrm{d} \over {\mathrm{d}q}}{u_{n,q}}\left( z \right)\mathrm{d}z} }.    \cr
\end{split}
\end{equation}
\end{widetext}

According to Zener's formulae~\cite{Zener1934,Niu1996}, the escaping rate of a particle from one energy band into its next energy band reads
\begin{equation}
\gamma = {{F_{m_f}d} \over {2\pi \hbar }}{{\mathop{\rm e}\nolimits} ^{ - {{md\Delta E_n^2} \over {4{\hbar ^2}F_{m_f}}}}},\label{gamma}
\end{equation}
where $\Delta {E_n}$ is the gap between $n$-th and $\left({n + 1}\right)$-th energy bands.
According to the escaping rate formula~(\ref{gamma}), if $F_{m_f}$ is very small or $\Delta {E_n}$ is very large, we have the escaping rate $\gamma \rightarrow 0$.
Under these assumptions, we obtain the single-band Hamiltonian
\begin{eqnarray}
\label{equ:single-n}
  \hat H = && F_{m_f}d {\hat{\sigma} _z} \sum\limits_{n,j,s}^{} {j \left|{n,j,s}\right\rangle \left\langle {n,j,s} \right|} \nonumber \\
  && + \sum\limits_{\left\langle {j,k} \right\rangle ,s}^{} {t_{j,k}^n\left| {n,j,s} \right\rangle \left\langle {n,k,s} \right|} \nonumber \\
  && + {{\hbar \Omega } \over 2}\sum\limits_{j}^{} \left({\left| {n,j,\uparrow} \right\rangle \left\langle {n,j,\downarrow} \right|} + \mathrm{h.c.} \right),
\end{eqnarray}
for the $n$-th energy band.
If $t_{j,j + 1}^n \gg t_{j,k}^n$ for ${\left|{j - k}\right| \ge 2}$, i.e., the long-range hopping is much smaller than the nearest neighbor hopping, one can use the tight-binding approximation.

For the lowest band, denoting $\left|{n=1,j,s}\right\rangle\equiv\left|{j,s}\right\rangle$, we have the following tight-binding Hamiltonian,
\begin{eqnarray}
\label{equ:single}
  \hat H = && F_{m_f}d {\hat{\sigma} _z} \sum\limits_{j,s}^{} {j\left| {j, s} \right\rangle \left\langle {j,s} \right|} + \varepsilon \sum\limits_{j,s}^{} {\left| {j,s} \right\rangle \left\langle {j,s} \right|} \nonumber \\
  && + \Delta\sum\limits_{j,s}^{} \left( {\left| {j,s} \right\rangle \left\langle {j+1,s} \right|}  + \mathrm{h.c.} \right) \nonumber \\
  && + {{\hbar \Omega } \over 2}\sum\limits_{j}^{} \left({\left| {j,\uparrow} \right\rangle \left\langle {j,\downarrow} \right|} + \mathrm{h.c.} \right).
\end{eqnarray}
Here, the on-site energy $\varepsilon=t_{j,j}^{n=1}$  and  the hopping strength $\Delta = t_{j,j + 1}^{n=1}$ are independent of the lattice site number $j$.

\end{document}